\documentclass[12pt]{article}

\usepackage{axodraw}

\makeatletter

\def\paragraph{\@startsection{paragraph}{4}{\z@}{+2.00ex plus
 +1ex minus +.2ex}{1.5ex plus .2ex}{\it\normalsize}}

\def\section{\@startsection {section}{1}{\z@}{+3.0ex plus +1ex minus
  +.2ex}{2.3ex plus .2ex}{\normalsize\bf\boldmath}}
\def\subsection{\@startsection{subsection}{2}{\z@}{+2.5ex plus +1ex
minus +.2ex}{1.5ex plus .2ex}{\normalsize\bf\boldmath}}
\def\subsubsection{\@startsection{subsubsection}{3}{\z@}{+3.25ex plus
 +1ex minus +.2ex}{1.5ex plus .2ex}{\normalsize\it}}

\expandafter\ifx\csname mathrm\endcsname\relax\def\mathrm#1{{\rm #1}}\fi

\newcounter{saveeqn}

\@addtoreset{equation}{section}

\newcount\@tempcntc
\def\@citex[#1]#2{\if@filesw\immediate\write\@auxout{\string\citation{#2}}\fi
  \@tempcnta\z@\@tempcntb\m@ne\def\@citea{}\@cite{\@for\@citeb:=#2\do
    {\@ifundefined
       {b@\@citeb}{\@citeo\@tempcntb\m@ne\@citea
        \def\@citea{,\penalty\@m\ }{\bf ?}\@warning
       {Citation `\@citeb' on page \thepage \space undefined}}%
    {\setbox\z@\hbox{\global\@tempcntc0\csname
b@\@citeb\endcsname\relax}%
     \ifnum\@tempcntc=\z@ \@citeo\@tempcntb\m@ne
       \@citea\def\@citea{,\penalty\@m}
       \hbox{\csname b@\@citeb\endcsname}%
     \else
      \advance\@tempcntb\@ne
      \ifnum\@tempcntb=\@tempcntc
      \else\advance\@tempcntb\m@ne\@citeo
      \@tempcnta\@tempcntc\@tempcntb\@tempcntc\fi\fi}}\@citeo}{#1}}

\def\@citeo{\ifnum\@tempcnta>\@tempcntb\else\@citea
  \def\@citea{,\penalty\@m}%
  \ifnum\@tempcnta=\@tempcntb\the\@tempcnta\else
   {\advance\@tempcnta\@ne\ifnum\@tempcnta=\@tempcntb \else
\def\@citea{--}\fi
    \advance\@tempcnta\m@ne\the\@tempcnta\@citea\the\@tempcntb}\fi\fi}

\def\nl{\nonumber\\}

\newcommand{\lsim}
{\mathrel{\raisebox{-.3em}{$\stackrel{\displaystyle <}{\sim}$}}}
\newcommand{\gsim}
{\mathrel{\raisebox{-.3em}{$\stackrel{\displaystyle >}{\sim}$}}}
\def\asymp#1%
{\mathrel{\raisebox{-.4em}{$\widetilde{\scriptstyle #1}$}}}

\def\Nequal#1%
{\mathrel{\raisebox{-.5em}{$\stackrel{=}{\scriptstyle\rm#1}$}}}
\newcommand{\dsl}[1]{\not \hspace{-0.7mm}#1}
\def\dsl{\mathpalette\make@slash}
\def\make@slash#1#2{\setbox\z@\hbox{$#1#2$}%
  \hbox to 0pt{\hss$#1/$\hss\kern-\wd0}\box0}

\def\beq{\begin{equation}}
\def\eeq{\end{equation}}
\def\beqar{\begin{eqnarray}}
\def\eeqar{\end{eqnarray}}
\def\barr#1{\begin{array}{#1}}
\def\earr{\end{array}}
\def\bfi{\begin{figure}}
\def\efi{\end{figure}}
\def\btab{\begin{table}}
\def\etab{\end{table}}
\def\bce{\begin{center}}
\def\ece{\end{center}}
\def\nn{\nonumber}
\def\disp{\displaystyle}
\def\text{\textstyle}

\def\al{\alpha}
\def\be{\beta}

\def\de{\delta}

\def\eps{\epsilon}

\def\si{\sigma}

\def\refeq#1{\mbox{(\ref{#1})}}

\def\reffi#1{\mbox{Figure~\ref{#1}}}
\def\reffis#1{\mbox{Figures~\ref{#1}}}
\def\refta#1{\mbox{Table~\ref{#1}}}

\def\refse#1{\mbox{Section~\ref{#1}}}

\def\citere#1{\mbox{Ref.~\cite{#1}}}
\def\citeres#1{\mbox{Refs.~\cite{#1}}}

\newcommand{\TeV}{\unskip\,\mathrm{TeV}}
\newcommand{\GeV}{\unskip\,\mathrm{GeV}}
\newcommand{\MeV}{\unskip\,\mathrm{MeV}}

\newcommand{\fba}{\unskip\,\mathrm{fb}}

\newcommand{\ri}{{\mathrm{i}}}
\newcommand{\rd}{{\mathrm{d}}}

\newcommand{\rE}{{\mathrm{E}}}

\newcommand{\Oa}{\mathswitch{{\cal{O}}(\alpha)}}

\newcommand{\M}{{\cal{M}}}

\def\mathswitchr#1{\relax\ifmmode{\mathrm{#1}}\else$\mathrm{#1}$\fi}
\newcommand{\Pf}{\mathswitch  f}

\newcommand{\PS}{\mathswitchr S}
\newcommand{\PP}{\mathswitchr P}
\newcommand{\PV}{\mathswitchr V}
\newcommand{\PW}{\mathswitchr W}
\newcommand{\Pw}{\mathswitchr w}
\newcommand{\PZ}{\mathswitchr Z}
\newcommand{\PA}{\mathswitchr A}
\newcommand{\Pg}{\mathswitchr g}
\newcommand{\PH}{\mathswitchr H}
\newcommand{\Ph}{\mathswitchr h}

\newcommand{\Pe}{\mathswitchr e}

\newcommand{\Ps}{\mathswitchr s}

\newcommand{\Pc}{\mathswitchr c}

\newcommand{\Pb}{\mathswitchr b}

\newcommand{\Pt}{\mathswitchr t}

\newcommand{\Pep}{\mathswitchr {e^+}}
\newcommand{\Pem}{\mathswitchr {e^-}}

\def\mathswitch#1{\relax\ifmmode#1\else$#1$\fi}
\newcommand{\Mf}{\mathswitch {m_\Pf}}

\newcommand{\MV}{\mathswitch {M_\PV}}
\newcommand{\MW}{\mathswitch {M_\PW}}
\newcommand{\MA}{\mathswitch {M_\PA}}
\newcommand{\MZ}{\mathswitch {M_\PZ}}
\newcommand{\Mh}{\mathswitch {M_\Ph}}
\newcommand{\MH}{\mathswitch {M_\PH}}
\newcommand{\MP}{\mathswitch {M_\PP}}
\newcommand{\MS}{\mathswitch {M_\PS}}

\newcommand{\Ms}{\mathswitch {m_\Ps}}
\newcommand{\Mc}{\mathswitch {m_\Pc}}
\newcommand{\Mb}{\mathswitch {m_\Pb}}
\newcommand{\Mt}{\mathswitch {m_\Pt}}
\newcommand{\GW}{\Gamma_{\PW}}
\newcommand{\GZ}{\Gamma_{\PZ}}

\newcommand{\Gh}{\Gamma_{\Ph}}
\newcommand{\GH}{\Gamma_{\PH}}
\newcommand{\GA}{\Gamma_{\PA}}

\newcommand{\sw}{\mathswitch {s_\Pw}}
\newcommand{\cw}{\mathswitch {c_\Pw}}

\newcommand{\GF}{\mathswitch {G_\mu}}

\newcommand{\QCD}{{\mathrm{QCD}}}

\newcommand{\LL}{\mathrm{LL}}

\newcommand{\soft}{{\mathrm{soft}}}

\newcommand{\CF}{{C_\mathrm{F}}}

\def\Li{\mathop{\mathrm{Li}_2}\nolimits}

\def\Re{\mathop{\mathrm{Re}}\nolimits}

\renewcommand{\O}{{\cal O}}

\def\draftdate{\relax}
\def\mda{\relax}
\def\mua{\relax}
\def\mla{\relax}
\def\draft{
\def\thtystars{******************************}
\def\sixtystars{\thtystars\thtystars}
\typeout{}
\typeout{\sixtystars**}
\typeout{* Draft mode!
         For final version remove \protect\draft\space in source file *}
\typeout{\sixtystars**}
\typeout{}
\def\draftdate{\today}
\def\mua{\marginpar[\boldmath\hfil$\uparrow$]%
                   {\boldmath$\uparrow$\hfil}%
                    \typeout{marginpar: $\uparrow$}\ignorespaces}
\def\mda{\marginpar[\boldmath\hfil$\downarrow$]%
                   {\boldmath$\downarrow$\hfil}%
                    \typeout{marginpar: $\downarrow$}\ignorespaces}
\def\mla{\marginpar[\boldmath\hfil$\rightarrow$]%
                   {\boldmath$\leftarrow $\hfil}%
                    \typeout{marginpar: $\leftrightarrow$}\ignorespaces}
\def\Mua{\marginpar[\boldmath\hfil$\Uparrow$]%
                   {\boldmath$\Uparrow$\hfil}%
                    \typeout{marginpar: $\uparrow$}\ignorespaces}
\def\Mda{\marginpar[\boldmath\hfil$\Downarrow$]%
                   {\boldmath$\Downarrow$\hfil}%
                    \typeout{marginpar: $\downarrow$}\ignorespaces}
\def\Mla{\marginpar[\boldmath\hfil$\Rightarrow$]%
                   {\boldmath$\Leftarrow $\hfil}%
                    \typeout{marginpar: $\leftrightarrow$}\ignorespaces}
\overfullrule 5pt
\oddsidemargin -15mm
\marginparwidth 29mm
}

\def\stars{\strut\leaders\hbox{*}\hfill\strut}
\def\starline{\hfil\strut\hfil\hbox to \textwidth {\stars}\hfil}

\begin{document}
\thispagestyle{empty}
\def\thefootnote{\fnsymbol{footnote}}
\setcounter{footnote}{1}
\null
\draftdate\hfill DESY 02-024 
\\
\strut\hfill PSI-PR-02-02\\
\strut\hfill hep-ph/0203120 
\vfill
\begin{center}
{\large \bf\boldmath
Photonic and QCD radiative corrections to
\\[.5em]
Higgs-boson production in $\mu^+\mu^-\to f\bar f$ 
\par} \vskip 2em
\vspace{1cm}
{\large
{\sc Stefan Dittmaier$^1$%
\footnote{Heisenberg fellow of the Deutsche Forschungsgemeinschaft}
and Andreas Kaiser$^2$} } 
\\[.5cm]
$^1$ {\it Deutsches Elektronen-Synchrotron DESY \\
D-22603 Hamburg, Germany}
\\[0.3cm]
$^2$ {\it Paul Scherrer Institut\\
CH-5232 Villigen PSI, Switzerland} 
\par 
\end{center}\par
\vfill
\vskip 2.0cm {\bf Abstract:} \par 
The photonic and QCD radiative corrections at next-to-leading order
are calculated for fermion-pair production at muon colliders, 
maintaining the full mass dependence and helicity information of
the muons and the produced fermions. Higher-order effects of 
initial-state radiation are included at the leading logarithmic
level. In the calculation particular attention is paid to the issue 
of gauge invariance in the vicinity of resonances. The most 
important corrections are presented in analytical form. 
The detailed numerical discussion concentrates on the corrections
to the ($s$-channel) Higgs-boson resonances in the Standard Model 
and its minimal supersymmetric extension. 
The results show that photonic initial-
and QCD final-state corrections are very important in a precision study of
Higgs resonances, but that (photonic) initial-final interferences are widely
suppressed and only modify the non-resonant background.
\par
\vskip 1cm
\noindent
March 2002   
\null
\setcounter{page}{0}
\clearpage
\def\thefootnote{\arabic{footnote}}
\setcounter{footnote}{0}

\section{Introduction}
\label{se:intro}

The search for Higgs bosons will be among the most prominent ambitions
in future high-energy collider experiments. Once one or even more
Higgs bosons are found at the 
Tevatron \cite{Carena:2000yx}, 
the LHC \cite{atlas_cms_tdrs}, 
or a future $\Pep\Pem$ collider \cite{Accomando:1998wt}, 
it is necessary to determine the properties of these
particles, such as their masses, decay widths, and couplings, in order 
to reconstruct the scalar sector of the underlying spontaneously broken
gauge theory. Current electroweak precision data 
\cite{Group:2001ix} and the direct search for a Higgs boson at LEP2
\cite{unknown:2001xw}
constrain the mass $\MH$ of the Standard Model (SM) Higgs boson
to $\MH<222\GeV$ and $\MH>114\GeV$ at the 95\% C.L., respectively.
In the Minimal Supersymmetric extensions of the SM (MSSM) the lower
bound for the light Higgs boson h is about $\Mh>91\GeV$ 
\cite{unknown:2001xx}, while an upper bound of $\Mh\lsim 130\GeV$ results from 
theoretical constraints (see e.g.\ \citeres{Carena:1996wu,Carena:2000dp} 
and references therein). 
Thus, it seems very promising to search for
a Higgs boson with mass $\lsim 140\GeV$. 
Since the decay channel into 
on-shell W and Z bosons is not yet 
open in this mass range, such a Higgs boson has a very narrow width,
rendering its {\it direct} measurement in a resonance distribution
impossible at hadron and $\Pep\Pem$ colliders.

A muon collider provides the unique opportunity to investigate
Higgs-boson resonances in the $s$-channel within the clean environment
of a lepton collider, since the Yukawa coupling of Higgs bosons to
the initial-state muons is about 200 times larger than for $\Pe^\pm$,
resulting in a relative enhancement by a factor 40000.
Studies \cite{Raja:1998ip}
of the experimental feasibility of a muon collider show that the
beam energy resolution can even be small enough to resolve a Higgs
resonance with a width of some MeV, as it for instance appears
for a SM-like Higgs boson with mass $\lsim 140\GeV$.
To be specific, for a SM Higgs boson of $\MH\sim 110\GeV$ the relative
accuracies achievable by scanning the $\mu^+\mu^-\to\PH\to\Pb\bar\Pb$
resonance with a luminosity of $0.2\fba^{-1}$ was 
estimated in \citeres{Barger:1995hr,Barger:1996jm,Barger:2001mi}
to be $\sim 1-3\times 10^{-6}$ and $0.2$ for the mass and total width,
respectively. Moreover, observing the reaction 
$\mu^+\mu^-\to\PH\to f\bar f$, is complementary to Higgs production
at $\Pep\Pem$ machines where the Higgs boson is produced via its
couplings to Z or W~bosons. The high accuracy at a muon collider
also allows for a better discrimination between
a SM and a SM-like Higgs boson in extended models like the MSSM
as compared to hadron or $\Pep\Pem$ colliders.
Furthermore, 
the existence of a SM-like Higgs boson h in the MSSM means
that the two other MSSM Higgs bosons A and H are heavy; in this case
a muon collider also proves to be the most promising machine to
study the properties of these Higgs bosons. For instance, 
measuring the mass difference $\MH-\MA$ yields an important check
of the Higgs mass relations \cite{Carena:1996wu,Carena:2000dp}
in the MSSM, as pointed out in \citere{Berger:2001et}.
More information on the Higgs phenomenology at muon colliders
can be found in the literature
\cite{Raja:1998ip,Barger:1995hr,Barger:1996jm,Barger:2001mi,Berger:2001et,Grzadkowski:1995rx,Barger:1999tj,Atwood:1995ej,Autin:1999ci}
including in particular studies
of CP violation \cite{Grzadkowski:1995rx},
Higgs decays into $\tau^+\tau^-$ pairs \cite{Barger:1999tj},
and flavour-changing neutral currents~\cite{Atwood:1995ej}.
In this context, the possibility to produce polarized muon beams
proved to be very useful.

In this paper we calculate the photonic and QCD corrections
to the fermion-pair production processes $\mu^+\mu^-\to f\bar f$
in the SM and the MSSM.
These corrections represent an important ingredient in a precision
calculation of the Higgs resonances in these channels.
In particular, photonic initial-state radiation (ISR)
distorts the resonance shape, as it is, e.g., well known 
from studies of the Z~resonance in $\Pep\Pem$ annihilation
(see e.g.\ \citere{LEP1} and references therein).
In contrast to the $\Pep\Pem$ case, however, we have to
include the finite-mass effects of both initial-state
and final-state fermions, since Higgs production proceeds
via the so-called spin-0 channel which is mass suppressed.
In detail, we calculate the full photonic ${\cal O}(\alpha)$
correction, which decomposes into ISR, final-state radiation
(FSR) and initial-final interferences, and include the leading
ISR effects beyond ${\cal O}(\alpha)$ in the structure-function
approach \cite{sf}. 
In some existing studies ISR effects were already included
at the leading logarithmic level as described in the appendix of
\citere{Barger:1996jm}; however, non-leading corrections have not yet
been presented elsewhere. These effects are of particular importance
if polarized muon beams are considered, since the leading logarithmic
approach 
ignores a possible spin
flip of radiating muons, which mixes the spin-0 and spin-1
channels.
QCD corrections are evaluated in
next-to-leading order, i.e.\ in ${\cal O}(\alpha_{\mathrm{s}})$,
also taking into account the full mass dependence. For
b~quarks in the final state the running $\overline{\mathrm{MS}}$ 
mass is introduced.

In our discussion of numerical results we mainly focus on
$\Pb\bar\Pb$ final states and consider SM Higgs-boson resonances
for $\MH=115\GeV$ and $150\GeV$. In the MSSM we discuss two
interesting scenarios, one where the H/A resonances lie on top
of each other and one where they are sufficiently separated in mass.
In the latter case, the masses are chosen large enough to enable
an inclusion of $\Pt\bar\Pt$ production. 
Cross sections for polarized muons are only briefly discussed,
although our calculation fully supports helicity eigenstates%
\footnote{The generalization to other polarization states, such as
transversely polarized beams, is straightforward.}
of the external fermions.

The non-photonic, i.e.\ purely weak,
electroweak radiative corrections of $\O(\GF)$
are not considered in this paper for the following reasons. 
It is known (see e.g.\ \citere{LEP1}) that photonic corrections, 
in particular ISR effects, comprise a very important gauge-invariant 
class of radiative corrections to resonance processes of neutral particles.
Another important source of corrections are renormalization effects,
such as the running of the electromagnetic coupling $\alpha$.
The latter universal effect 
is included by appropriately choosing the input value of $\alpha$.
Moreover, a fully convincing calculation of the non-photonic $\O(\GF)$
corrections is intrinsically connected to the Dyson summation of all
relevant propagator corrections without losing gauge invariance.
We discuss this issue in detail for the photonic and QCD corrections
and give an outlook to the non-photonic corrections, where the
problem is more involved.
In summary, our results can be viewed as a first step towards a
precision calculation, and the weak corrections are part in one of
the next steps. If one is only interested in a precise prediction
of the Higgs resonance shapes, a reasonable approach would be to
include the weak corrections to the Higgs Yukawa couplings on
the Higgs resonances in the framework of effective couplings.
This procedure maintains gauge invariance and the effective couplings
can be easily combined with the photonic and QCD corrections discussed 
here. The most important weak corrections to the Higgs Yukawa couplings
in the MSSM have already been discussed in \citere{Pierce:1996zz}.

The paper is organized as follows:
In \refse{se:lo} we set our conventions and calculate the lowest-order
cross sections. Section~\ref{se:gaugeinv} contains our discussion of
gauge invariance. The photonic corrections are treated in
\refse{se:emrcs}, where our explicit analytical results for the ISR and 
FSR effects can be found. The QCD corrections are deduced from the
photonic FSR results in \refse{se:qcdrcs}. Section~\ref{se:num}
provides a detailed discussion of numerical results for the various
corrections near the Higgs resonances.
Our conclusions are given in \refse{se:concl}, and the Appendix
provides some auxiliary formulas.

\section{Notation and lowest-order cross sections}
\label{se:lo}

\begin{figure}
\setlength{\unitlength}{1pt}
\centerline{
\begin{picture}(120,85)(0,0)
\ArrowLine(10,70)(40,40)
\ArrowLine(40,40)(10,10)
\Photon   (40,40)(80,40){2}{4}
\ArrowLine(80,40)(110,70)
\ArrowLine(110,10)(80,40)
\Vertex   (40,40){2.0}
\Vertex   (80,40){2.0}
\Text     ( -8,70)[l]{$\mu^-$}
\Text     ( -8,10)[l]{$\mu^+$}
\Text     (115,10)[l]{$\bar f$}
\Text     (115,70)[l]{$f$}
\Text     ( 50,28)[l]{$\gamma,\PZ$}
\end{picture}
\hspace*{2em}
\begin{picture}(120,85)(0,0)
\ArrowLine(10,70)(40,40)
\ArrowLine(40,40)(10,10)
\DashLine(40,40)(80,40){5}
\ArrowLine(80,40)(110,70)
\ArrowLine(110,10)(80,40)
\Vertex   (40,40){2.0}
\Vertex   (80,40){2.0}
\Text     ( -8,70)[l]{$\mu^-$}
\Text     ( -8,10)[l]{$\mu^+$}
\Text     (115,10)[l]{$\bar f$}
\Text     (115,70)[l]{$f$}
\Text     ( 50,28)[l]{$\chi,\varphi$}
\end{picture} }
\caption{Lowest-order diagrams for $\mu^-\mu^+\to f\bar f$ with $f\ne\mu^-$}
\label{fi:diags_mumuff}
\end{figure}
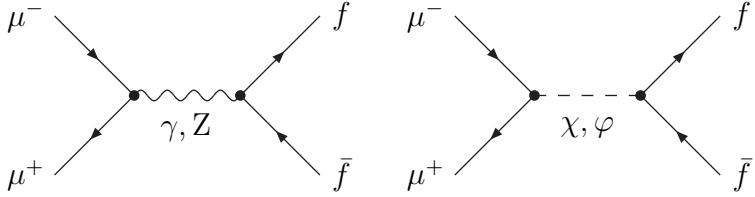

We consider the process
\beq
\mu^-(p,\sigma) + \mu^+(p',\sigma') \;\longrightarrow\;
f(q,\tau) + \bar f(q',\tau'), \qquad f\ne\mu^-,
\label{eq:mumuff}
\eeq
where the momenta $p$,\dots and helicities $\sigma$,\dots of the
muons and fermions $f$ are given in parentheses. 
The heliticities take the values $\sigma=\pm\frac{1}{2}$, etc., but
we will also take their sign to indicate the helicity.
In lowest order the diagrams of \reffi{fi:diags_mumuff} contribute
to the process \refeq{eq:mumuff}. 
In order to treat the SM and MSSM
in one go, we have abbreviated the Higgs bosons by $\varphi$,
where $\varphi=\PH$ in the SM and $\varphi=\Ph,\PH,\PA$ in the MSSM.
The field $\chi$ denotes the would-be Goldstone partner of the
Z~boson; in the unitary gauge the diagram involving $\chi$ exchange
is absent.

In the following we denote the vector and axial-vector couplings of the neutral 
gauge bosons to the fermions by
\beq
v_{\gamma f} = -Q_f, \qquad
a_{\gamma f} = 0, \qquad
v_{\PZ f} = \frac{\sw}{\cw}Q_f - \frac{I_{\Pw,f}}{2\cw\sw}, \qquad
a_{\PZ f} = -\frac{I_{\Pw,f}}{2\cw\sw},
\label{eq:gaugecoupls}
\eeq
where $I_{\Pw,f}=\pm\frac{1}{2}$ is the weak isospin of the left-handed
part of the fermion field $f$, and $Q_f$ is its electric charge.
The corresponding Feynman rule for the $V_\mu f\bar f$ vertex is given by
$\ri e\gamma_\mu (v_{\PV f}-a_{\PV f}\gamma_5)$,
where $e$ is the electric unit charge.
In these couplings the parameters
$\sw$ and $\cw$ are functions of the weak mixing angle $\theta_\Pw$,
\beq
\sin^2\theta_\Pw = \sw^2 = 1-\cw^2, \qquad \cw=\frac{\MW}{\MZ},
\label{eq:sw}
\eeq
and thus fixed by the ratio of the gauge-boson masses $\MW$ and $\MZ$.
For the vertices $Sf\bar f$ and $Pf\bar f$ of fermions with the neutral
scalar and pseudo-scalar fields S and P we generically 
write the Feynman rules
as $\ri eg_{\PS f}$ and $eg_{\PP f}\gamma_5$, respectively.
For the pseudo-scalar would-be Goldstone field $\chi$ the coupling factor
reads
\beq
g_{\chi f} = -\frac{I_{\Pw,f}}{\sw\cw}\frac{\Mf}{\hat\MZ},
\label{eq:cff}
\eeq
where $m_f$ is the mass of the fermion $f$. Here we write $\hat\MZ$
for the Z-boson mass instead of $\MZ$ for later convenience;
the definition of $\hat\MZ$ and its relation to $\MZ$ 
will be discussed in \refse{se:gaugeinv} in detail.
While the $\chi f\bar f$ coupling is equal in the SM and the MSSM, the
couplings of the fermions to the Higgs bosons, of course, differ in these
models. We follow the conventions of \citere{Gunion:1984yn} also
for the SM and define
\beq
\begin{array}[b]{rrlrl}
\mbox{SM:} \qquad &
g_{\PH f} &= \disp -\frac{1}{2\cw\sw}\frac{\Mf}{\MZ},
\\[1.3em]
\mbox{MSSM:} \qquad &
  g_{\Ph u} &= \disp 
	-\frac{1}{2\cw\sw}\frac{m_u}{\MZ}\frac{\cos\alpha}{\sin\beta}, \qquad &
  g_{\Ph d} &= \disp 
	+\frac{1}{2\cw\sw}\frac{m_d}{\MZ}\frac{\sin\alpha}{\cos\beta}, 
\\[.8em]
& g_{\PH u} &= \disp 
	-\frac{1}{2\cw\sw}\frac{m_u}{\MZ}\frac{\sin\alpha}{\sin\beta}, \qquad &
  g_{\PH d} &= \disp 
	-\frac{1}{2\cw\sw}\frac{m_d}{\MZ}\frac{\cos\alpha}{\cos\beta}, 
\\[.8em]
& g_{\PA u} &= \disp -\frac{1}{2\cw\sw}\frac{m_u}{\MZ}\cot\beta, \qquad &
  g_{\PA d} &= \disp -\frac{1}{2\cw\sw}\frac{m_d}{\MZ}\tan\beta,
\end{array}
\label{eq:higgscoupls}
\eeq
where $u$ and $d$ refer to up- and down-type fermions, respectively.
In the SM the Yukawa coupling $eg_{\PH f}$ is entirely determined
by the ratio $m_f/v$, where $v=1/\sqrt{\sqrt{2}\GF}$ is the 
vacuum expectation value of the Higgs field.
In the MSSM the couplings additionally depend on the angles $\alpha$
and $\beta$, which are related to the masses of the Higgs bosons.
We follow the usual convention to use the mass $\MA$ and 
$\tan\beta=v_2/v_1$ as independent input parameters, where $v_{1,2}$ 
are the vacuum expectation values of the two Higgs doublets in the
MSSM. The angle $\alpha$ as well as the other Higgs masses $\Mh$ and $\MH$
are then derived from $\MA$, $\tan\beta$, and the parameters of the
gauge sector; to this end we employ the program {\sl HDECAY}
\cite{Djouadi:1997yw} which calculates $\alpha$, the dependent Higgs masses,
and the Higgs decay widths $\Gamma_\varphi$, thereby including
also some leading radiative corrections \cite{Carena:1996wu}
to these relations.

Sometimes it is more convenient to decompose the
fermionic couplings into their right- and left-handed parts, which
are proportional to the chirality projectors $\omega_\pm=(1\pm\gamma_5)/2$,
instead of using vector, axial-vector, scalar, and pseudo-scalar couplings.
Therefore, we introduce the right- and left-handed coupling
factors $g^\pm_{\dots}$ for the various couplings introduced above,
\beq
g_{\PV f}^\pm = v_{\PV f} \mp a_{\PV f}, \qquad
g_{\PS f}^\pm = g_{\PS f}, \qquad
g_{\PP f}^\pm = \mp \ri g_{\PP f},
\label{eq:chiralcoupls}
\eeq
so that the Feynman rule for the $V_\mu f\bar f$ vertex is given by
$\ri e\gamma_\mu (g_{\PV f}^+\omega_+ + g_{\PV f}^-\omega_-)$
and the rules for the scalar or pseudo-scalar field $\varphi=\PS,\PP$
generically read $\ri e (g_{\varphi f}^+\omega_+ + g_{\varphi f}^-\omega_-)$.

The Mandelstam variables for the process \refeq{eq:mumuff} 
are defined as usual,
\beq
s=(p+p')^2, \qquad 
t=(p-q)^2, \qquad
u=(p-q')^2.
\eeq
In the calculation of the tree-level amplitude $\M_0$ and of the
one-loop amplitude $\M_1$, which is described in the next section, 
we separate the fermion spinor chains by defining the following
standard matrix elements (SME),
\beqar
\M^{\{v,a\}\{v,a\}}_1 &=&
\Big[\bar v_{\mu^+} \gamma_\rho \{1,\gamma_5\} u_{\mu^-} \Big] \,
\Big[\bar u_{f} \gamma^\rho \{1,\gamma_5\} v_{\bar f} \Big],
\nn\\
\M^{\{v,a\}\{v,a\}}_2 &=&
\Big[\bar v_{\mu^+} \{1,\gamma_5\} u_{\mu^-} \Big] \,
\Big[\bar u_{f} \{1,\gamma_5\} v_{\bar f} \Big],
\nn\\
\M^{\{v,a\}\{v,a\}}_3 &=&
\Big[\bar v_{\mu^+} \dsl{q} \{1,\gamma_5\} u_{\mu^-} \Big] \,
\Big[\bar u_{f} \{1,\gamma_5\} v_{\bar f} \Big],
\nn\\
\M^{\{v,a\}\{v,a\}}_4 &=&
\Big[\bar v_{\mu^+} \{1,\gamma_5\} u_{\mu^-} \Big] \,
\Big[\bar u_{f} \dsl{p} \{1,\gamma_5\} v_{\bar f} \Big],
\label{eq:SME}
\eeqar
with an obvious notation for the Dirac spinors $\bar v_{\mu^+}$, etc.
The labels $v$ and $a$ in the SME refer to the insertions
of 1 or $\gamma_5$ in the two spinor chains; for instance we have
$\M^{va}_2=[\bar v_{\mu^+} u_{\mu_-}] [\bar u_{f} \gamma_5 v_{\bar f}]$.
The explicit expressions of the SME for all possible helicity
configurations can be found in the Appendix.
The SME represent 16 independent functions which form a basis for the
tree-level and one-loop amplitudes, i.e.\ these amplitudes admit the
representation
\beq
\M_n = \sum_{i=1}^4 \sum_{x,y=v,a} F^{xy}_{n,i}(s,t) \M^{xy}_i, \qquad n=0,1,
\eeq
with Lorentz invariant functions $F^{xy}_{n,i}(s,t)$.
The lowest-order amplitude reads
\beq
\M_0 = \sum_{V=\gamma,\PZ} \M_{\PV,0}
+\sum_{\PP=\chi,\PA} \M_{\PP,0} +\sum_{\PS=\Ph,\PH}  \M_{\PS,0}
\label{eq:M0}
\eeq
with
\beqar
\M_{\PV,0} &=& \frac{e^2}{s-\hat\MV^2}
\left(  v_{\PV\mu} v_{\PV f} \M^{vv}_1 - v_{\PV\mu} a_{\PV f} \M^{va}_1
      - a_{\PV\mu} v_{\PV f} \M^{av}_1 + a_{\PV\mu} a_{\PV f} \M^{aa}_1 \right),
\nn\\
\M_{\PP,0} &=& \frac{e^2 g_{\PP\mu}g_{\PP f}}{s-\hat\MP^2} \, \M^{aa}_2,
\nn\\
\M_{\PS,0} &=& -\frac{e^2 g_{\PS\mu}g_{\PS f}}{s-\hat\MS^2} \, \M^{vv}_2,
\label{eq:MVPS0}
\eeqar
where it is understood that the sums over $\PP$ and $\PS$ extend only 
over $\chi$ and $\PH$, respectively, in the case of the SM. 
In a pure lowest-order prediction the propagator denominators $(s-M^2)$
lead to infinities in the resonance point $s=M^2$. A perturbative
description of these resonances requires a Dyson summation of
propagator corrections that regularize the unphysical divergence.
Here we signal the propagator modification, which is explained in the 
next section in detail, by putting carets over the masses in the propagators.
For the photon we, of course, have $\hat M_\gamma=0$.

Ignoring the Dyson summation for the moment, in Eq.~\refeq{eq:M0}
the contribution of $\PP=\chi$ results from the sum of 
$\chi$ exchange and of the $k_\mu k_\nu$ part of Z-boson propagator
in a general $R_\xi$ gauge, with $k$ denoting the transferred momentum. 
The unitary gauge and the `t~Hooft--Feynman 
gauge represent two interesting special cases. In the unitary gauge
the $\chi$ diagram is absent and the $\PP=\chi$ part is 
completely due to the $k_\mu k_\nu$ part of the Z propagator; in the
`t~Hooft--Feynman it is the other way round.

Finally, the lowest-order cross section reads
\beq
\sigma_0 = \frac{N_{\mathrm{c},f}\beta_f}{64\pi^2 s \beta_\mu} 
\int\rd\Omega_f \sum_{\sigma,\sigma',\tau,\tau'}
\frac{1}{4}(1+2P_-\sigma)(1+2P_+\sigma') \, 
|\M_0(\sigma,\sigma',\tau,\tau')|^2,
\label{eq:sigma0}
\eeq
where 
\beq
\beta_\mu = \sqrt{1-\frac{4m_\mu^2}{s}}, \qquad
\beta_f   = \sqrt{1-\frac{4m_f^2}{s}},
\eeq
are the velocities of the muons and produced fermions in the centre-of-mass
(CM) frame, where also the solid angle $\Omega_f$ of $f$ is defined,
and $N_{\mathrm{c},f}$ is the number of colours of $f$,
which is 1 if $f$ is a lepton and 3 if $f$ is a quark.
The degrees of polarization in the $\mu^\pm$ beams are denoted by 
$P_\pm$ and range from $-1$ to $+1$.

\section{The issue of gauge invariance}
\label{se:gaugeinv}

\subsection{Preliminaries}

We first consider the case of a single resonance appearing in a
propagator with momentum transfer $k$ and (renormalized)
mass $M$. In lowest order the resonance
appears for $k^2\to M^2$ where the lowest-order propagator has a pole, and
a proper description of the resonance requires a Dyson summation of 
the self-energy corrections to the resonant propagator. 
After this summation, in the vicinity of the resonance the amplitude 
$\M$ behaves as
\beq
\M \;\sim\; \frac{R}{k^2-M^2+\Sigma(k^2)} \,+\,
\mbox{non-resonant terms,}
\label{eq:Mres}
\eeq
where $\Sigma(k^2)$ denotes the self-energy of the propagating 
particle. The resonance appears for a value of $k^2$ where
$k^2-M^2+\Re\{\Sigma(k^2)\}=0$.
Since $\Sigma(k^2)$ develops a non-vanishing imaginary part in this 
resonance region, the propagator pole is shifted away from the real axis
into the complex $k^2$ plane. We denote the complex location of this
pole by $\bar M^2-\ri\bar M\bar\Gamma$ in the following.
The real quantities $\bar M$ and $\bar\Gamma$ provide a consistent 
definition of the particle's mass and width, respectively. In particular,
for gauge theories it was shown \cite{Gambino:1999ai}
that this definition is related to the bare parameters of the theory
in a gauge-invariant way.
Making use of $\bar M$ and $\bar\Gamma$, the naive propagator modification
\beq
\frac{1}{k^2-M^2} \;\to\; \frac{1}{k^2-\bar M^2+\ri\bar M\bar\Gamma},
\label{eq:fws}
\eeq
which is known as {\it fixed-width scheme},
yields a possible description of the resonance in lowest order.

Another definition of the particle's mass is provided by the zero
of the real part of the resonance denominator in Eq.~\refeq{eq:Mres},
known as on-shell definition, which we identify with $M$ in the following.
This definition was, for instance, adopted for the Z-boson mass at LEP1.
The quantities $M$ and $\bar M$ coincide
up to one-loop order, but differ at two loops and beyond.
Making use of the usual renormalization conditions in the on-shell scheme,
$\Re\{\Sigma(M^2)\}=0$ and $\Re\{\Sigma'(M^2)\}=0$, the resonance
behaviour of the propagator in Eq.~\refeq{eq:Mres} is obtained from 
the lowest order by the substitution
\beq
\frac{1}{k^2-M^2} \;\to\; \frac{1}{k^2-M^2+\ri M\Gamma(k^2)}.
\label{eq:rws}
\eeq
In the approximation of massless decay products 
the running width $\Gamma(k^2)$ is related to the on-shell decay 
width $\Gamma$ by $\Gamma(k^2)=\Gamma\times k^2/M^2\,\theta(k^2)$.
The parametrization \refeq{eq:rws} is known as {\it running-width
scheme}. 
It can be checked easily that the parameter sets
$(\bar M,\bar\Gamma)$ and $(M,\Gamma)$, when extracted upon fitting the 
parametrizations \refeq{eq:fws} and \refeq{eq:rws} to data, are
related by \cite{Bardin:1988xt}
\beqar
\bar M &=& M/\sqrt{1+\gamma^2}
= M-\frac{\Gamma^2}{2M} + \dots, \qquad \gamma=\Gamma/M,
\nn\\
\bar\Gamma &=& \Gamma/\sqrt{1+\gamma^2}
= \Gamma-\frac{\Gamma^3}{2M^2} + \dots,
\nn\\
\bar R &=& R/(1+\ri\gamma)
= R\left(1-\ri\frac{\Gamma}{M} + \dots \right),
\label{eq:massrel}
\eeqar
where we have also included the relation of the corresponding 
residues $\bar R$ and $R$.

At first sight it seems that the naive propagator modifications
\refeq{eq:fws} and \refeq{eq:rws}
yield reasonable descriptions of the resonance at least in lowest order.
However, this is not the case in general, since this modification 
potentially violates the gauge-invariance properties of the full
amplitude. Gauge invariance implies relations between self-energy, 
vertex, and box corrections (etc.) that in particular guarantee the
cancellation of gauge-fixing parameters within the amplitude.
These generalized Ward identities work order by order in perturbation
theory, i.e.\ they are jeopardized by any kind of partial inclusion
of higher-order corrections, such as the modification \refeq{eq:fws}
which resums part of the self-energy corrections.
Therefore, finite decay widths have to be introduced very carefully,
in particular if radiative corrections are taken into account,
which additionally complicate the situation. 

\subsection{Survey of photonic and QCD corrections --- 
gauge-invariant subsets of Feynman graphs}
\label{se:RCsurvey}

Before we explicitly describe our treatment of the resonances occurring in 
the process \refeq{eq:mumuff}, it is useful to identify gauge-invariant 
subsets of Feynman graphs within the inspected class of corrections.

At tree level (see \reffi{fi:diags_mumuff}) only Z-boson and $\chi$ exchange
diagrams are linked by gauge invariance, while each diagram with
photon or Higgs-boson exchange is gauge invariant in itself. 
This is also reflected by the fact that the Z- and Higgs-boson masses in the
propagators can be viewed as independent parameters in the lowest-order
diagrams. 
While this argument is trivially correct in the SM, where $\MH$ is a free
parameter, it works in the MSSM since 
the same diagrams appear, for instance, in a general
two-Higgs-doublet model, where the masses are completely unrelated.
Thus, the MSSM relations among the Higgs masses do not disturb this
gauge-invariance property of the (tree-level) diagrams.

Next we consider the photonic $\O(\alpha)$ corrections, which
are induced by one-loop photon exchange diagrams and real emission
of a single photon. This subclass of electroweak corrections is
gauge-invariant, since a consistent U(1)$_{\mathrm{elmg}}$ gauge
theory of the neutral bosons ($\gamma$, Z, and Higgs bosons) and all other 
fermions is conceivable, in which the process \refeq{eq:mumuff} proceeds
in the same way as in the SM or MSSM in lowest order. If W~bosons
were involved at tree level this would be not the case.
\begin{figure}
\setlength{\unitlength}{1pt}
\centerline{
\begin{picture}(120,85)(0,0)
\ArrowLine(10,70)(25,55)
\ArrowLine(25,55)(40,40)
\ArrowLine(40,40)(25,25)
\ArrowLine(25,25)(10,10)
\Photon   (40,40)(80,40){2}{4}
\ArrowLine(80,40)(110,70)
\ArrowLine(110,10)(80,40)
\Photon   (25,55)(25,25){2}{4}
\Vertex   (40,40){2.0}
\Vertex   (80,40){2.0}
\Vertex   (25,55){2.0}
\Vertex   (25,25){2.0}
\Text     ( -35,70)[l]{\bf (a)}
\Text     ( -8,70)[l]{$\mu^-$}
\Text     ( -8,10)[l]{$\mu^+$}
\Text     (115,10)[l]{$\bar f$}
\Text     (115,70)[l]{$f$}
\Text     ( 12,38)[l]{$\gamma$}
\Text     ( 50,28)[l]{$\gamma,\PZ$}
\end{picture}
\hspace*{2em}
\begin{picture}(120,85)(0,0)
\ArrowLine(10,70)(25,55)
\ArrowLine(25,55)(40,40)
\ArrowLine(40,40)(25,25)
\ArrowLine(25,25)(10,10)
\DashLine(40,40)(80,40){5}
\ArrowLine(80,40)(110,70)
\ArrowLine(110,10)(80,40)
\Photon   (25,55)(25,25){2}{4}
\Vertex   (40,40){2.0}
\Vertex   (80,40){2.0}
\Vertex   (25,55){2.0}
\Vertex   (25,25){2.0}
\Text     ( -8,70)[l]{$\mu^-$}
\Text     ( -8,10)[l]{$\mu^+$}
\Text     (115,10)[l]{$\bar f$}
\Text     (115,70)[l]{$f$}
\Text     ( 12,38)[l]{$\gamma$}
\Text     ( 50,28)[l]{$\chi,\varphi$}
\end{picture} }
\vspace*{.5em}
\centerline{
\begin{picture}(120,85)(0,0)
\ArrowLine(10,70)(40,40)
\ArrowLine(40,40)(10,10)
\Photon   (40,40)(80,40){2}{4}
\ArrowLine(80,40)( 95,55)
\ArrowLine(95,55)(110,70)
\ArrowLine(110,10)(95,25)
\ArrowLine( 95,25)(80,40)
\Photon   (95,55)(95,25){2}{4}
\Vertex   (40,40){2.0}
\Vertex   (80,40){2.0}
\Vertex   (95,55){2.0}
\Vertex   (95,25){2.0}
\Text     ( -35,70)[l]{\bf (b)}
\Text     ( -8,70)[l]{$\mu^-$}
\Text     ( -8,10)[l]{$\mu^+$}
\Text     (115,10)[l]{$\bar f$}
\Text     (115,70)[l]{$f$}
\Text     (102,38)[l]{$\gamma$}
\Text     ( 50,28)[l]{$\gamma,\PZ$}
\end{picture}
\hspace*{2em}
\begin{picture}(120,85)(0,0)
\ArrowLine(10,70)(40,40)
\ArrowLine(40,40)(10,10)
\DashLine(40,40)(80,40){5}
\ArrowLine(80,40)( 95,55)
\ArrowLine(95,55)(110,70)
\ArrowLine(110,10)(95,25)
\ArrowLine( 95,25)(80,40)
\Photon   (95,55)(95,25){2}{4}
\Vertex   (40,40){2.0}
\Vertex   (80,40){2.0}
\Vertex   (95,55){2.0}
\Vertex   (95,25){2.0}
\Text     ( -8,70)[l]{$\mu^-$}
\Text     ( -8,10)[l]{$\mu^+$}
\Text     (115,10)[l]{$\bar f$}
\Text     (115,70)[l]{$f$}
\Text     (102,38)[l]{$\gamma$}
\Text     ( 50,28)[l]{$\chi,\varphi$}
\end{picture} }
\caption{Diagrams for photonic vertex corrections to
$\mu^-\mu^+\to f\bar f$ contributing to (a) initial-state and (b) 
final-state corrections with $\varphi$ denoting any Higgs boson}
\label{fi:diags_vert}
%
\vspace*{2em}
\setlength{\unitlength}{1pt}
\centerline{
\begin{picture}(120,85)(0,0)
\ArrowLine(10,70)(40,65)
\ArrowLine(40,65)(40,15)
\ArrowLine(40,15)(10,10)
\Photon   (40,65)(80,65){2}{4}
\Photon   (40,15)(80,15){2}{4}
\ArrowLine(80,65)(110,70)
\ArrowLine(80,15)(80,65)
\ArrowLine(110,10)(80,15)
\Vertex   (40,15){2.0}
\Vertex   (80,15){2.0}
\Vertex   (40,65){2.0}
\Vertex   (80,65){2.0}
\Text     ( -8,70)[l]{$\mu^-$}
\Text     ( -8,10)[l]{$\mu^+$}
\Text     ( 27,38)[l]{$\mu$}
\Text     (115,10)[l]{$\bar f$}
\Text     (115,70)[l]{$f$}
\Text     ( 87,38)[l]{$f$}
\Text     ( 55,77)[l]{$V_1$}
\Text     ( 55,03)[l]{$V_2$}
\end{picture}
\hspace*{2em}
\begin{picture}(120,85)(0,0)
\ArrowLine(10,70)(40,65)
\ArrowLine(40,65)(40,15)
\ArrowLine(40,15)(10,10)
\Photon   (40,65)(80,65){2}{4}
\DashLine(40,15)(80,15){5}
\ArrowLine(80,65)(110,70)
\ArrowLine(80,15)(80,65)
\ArrowLine(110,10)(80,15)
\Vertex   (40,15){2.0}
\Vertex   (80,15){2.0}
\Vertex   (40,65){2.0}
\Vertex   (80,65){2.0}
\Text     ( -8,70)[l]{$\mu^-$}
\Text     ( -8,10)[l]{$\mu^+$}
\Text     ( 27,38)[l]{$\mu$}
\Text     (115,10)[l]{$\bar f$}
\Text     (115,70)[l]{$f$}
\Text     ( 87,38)[l]{$f$}
\Text     ( 57,77)[l]{$\gamma$}
\Text     ( 50,03)[l]{$\chi,\varphi$}
\end{picture} 
\hspace*{2em}
\begin{picture}(120,85)(0,0)
\ArrowLine(10,70)(40,65)
\ArrowLine(40,65)(40,15)
\ArrowLine(40,15)(10,10)
\Photon   (40,15)(80,15){2}{4}
\DashLine(40,65)(80,65){5}
\ArrowLine(80,65)(110,70)
\ArrowLine(80,15)(80,65)
\ArrowLine(110,10)(80,15)
\Vertex   (40,15){2.0}
\Vertex   (80,15){2.0}
\Vertex   (40,65){2.0}
\Vertex   (80,65){2.0}
\Text     ( -8,70)[l]{$\mu^-$}
\Text     ( -8,10)[l]{$\mu^+$}
\Text     ( 27,38)[l]{$\mu$}
\Text     (115,10)[l]{$\bar f$}
\Text     (115,70)[l]{$f$}
\Text     ( 87,38)[l]{$f$}
\Text     ( 57,03)[l]{$\gamma$}
\Text     ( 50,77)[l]{$\chi,\varphi$}
\end{picture} }
\vspace*{1em}
\centerline{
\begin{picture}(120,85)(0,0)
\ArrowLine(10,70)(40,65)
\ArrowLine(40,65)(40,15)
\ArrowLine(40,15)(10,10)
\Photon   (40,65)(80,15){2}{6}
\Photon   (40,15)(80,65){2}{6}
\ArrowLine(80,65)(110,70)
\ArrowLine(80,15)(80,65)
\ArrowLine(110,10)(80,15)
\Vertex   (40,15){2.0}
\Vertex   (80,15){2.0}
\Vertex   (40,65){2.0}
\Vertex   (80,65){2.0}
\Text     ( -8,70)[l]{$\mu^-$}
\Text     ( -8,10)[l]{$\mu^+$}
\Text     ( 27,38)[l]{$\mu$}
\Text     (115,10)[l]{$\bar f$}
\Text     (115,70)[l]{$f$}
\Text     ( 87,38)[l]{$f$}
\Text     ( 50,68)[l]{$V_1$}
\Text     ( 50,12)[l]{$V_2$}
\end{picture}
\hspace*{2em}
\begin{picture}(120,85)(0,0)
\ArrowLine(10,70)(40,65)
\ArrowLine(40,65)(40,15)
\ArrowLine(40,15)(10,10)
\Photon   (40,65)(80,15){2}{6}
\DashLine(40,15)(80,65){5}
\ArrowLine(80,65)(110,70)
\ArrowLine(80,15)(80,65)
\ArrowLine(110,10)(80,15)
\Vertex   (40,15){2.0}
\Vertex   (80,15){2.0}
\Vertex   (40,65){2.0}
\Vertex   (80,65){2.0}
\Text     ( -8,70)[l]{$\mu^-$}
\Text     ( -8,10)[l]{$\mu^+$}
\Text     ( 27,38)[l]{$\mu$}
\Text     (115,10)[l]{$\bar f$}
\Text     (115,70)[l]{$f$}
\Text     ( 87,38)[l]{$f$}
\Text     ( 50,68)[l]{$\gamma$}
\Text     ( 50,12)[l]{$\chi,\varphi$}
\end{picture} 
\hspace*{2em}
\begin{picture}(120,85)(0,0)
\ArrowLine(10,70)(40,65)
\ArrowLine(40,65)(40,15)
\ArrowLine(40,15)(10,10)
\Photon   (40,15)(80,65){2}{6}
\DashLine(40,65)(80,15){5}
\ArrowLine(80,65)(110,70)
\ArrowLine(80,15)(80,65)
\ArrowLine(110,10)(80,15)
\Vertex   (40,15){2.0}
\Vertex   (80,15){2.0}
\Vertex   (40,65){2.0}
\Vertex   (80,65){2.0}
\Text     ( -8,70)[l]{$\mu^-$}
\Text     ( -8,10)[l]{$\mu^+$}
\Text     ( 27,38)[l]{$\mu$}
\Text     (115,10)[l]{$\bar f$}
\Text     (115,70)[l]{$f$}
\Text     ( 87,38)[l]{$f$}
\Text     ( 50,12)[l]{$\gamma$}
\Text     ( 50,68)[l]{$\chi,\varphi$}
\end{picture} }
\caption{Photonic box diagrams for $\mu^-\mu^+\to f\bar f$ with
$V_1 V_2=\gamma\gamma,\gamma Z, Z\gamma$ and $\varphi$ denoting any Higgs boson}
\label{fi:diags_box}
%
\vspace*{2em}
\setlength{\unitlength}{1pt}
\centerline{
\begin{picture}(120,85)(0,0)
\ArrowLine(10,70)(25,55)
\ArrowLine(25,55)(40,40)
\ArrowLine(40,40)(10,10)
\Photon   (40,40)(80,40){2}{4}
\ArrowLine(80,40)(110,70)
\ArrowLine(110,10)(80,40)
\Photon   (25,55)(60,70){2}{4}
\Vertex   (40,40){2.0}
\Vertex   (80,40){2.0}
\Vertex   (25,55){2.0}
\Text     ( -8,70)[l]{$\mu^-$}
\Text     ( -8,10)[l]{$\mu^+$}
\Text     (115,10)[l]{$\bar f$}
\Text     (115,70)[l]{$f$}
\Text     ( 68,70)[l]{$\gamma$}
\Text     ( 50,28)[l]{$\gamma,\PZ$}
\end{picture}
\hspace*{2em}
\begin{picture}(150,85)(0,0)
\ArrowLine(10,70)(25,55)
\ArrowLine(25,55)(40,40)
\ArrowLine(40,40)(10,10)
\DashLine(40,40)(80,40){5}
\ArrowLine(80,40)(110,70)
\ArrowLine(110,10)(80,40)
\Photon   (25,55)(60,70){2}{4}
\Vertex   (40,40){2.0}
\Vertex   (80,40){2.0}
\Vertex   (25,55){2.0}
\Text     ( -8,70)[l]{$\mu^-$}
\Text     ( -8,10)[l]{$\mu^+$}
\Text     (115,10)[l]{$\bar f$}
\Text     (115,70)[l]{$f$}
\Text     ( 68,70)[l]{$\gamma$}
\Text     ( 50,28)[l]{$\chi,\varphi$}
\Text     (135,40)[l]{etc.}
\end{picture} }
\caption{Lowest-order diagrams for $\mu^-\mu^+\to f\bar f\gamma$ with 
$\varphi$ denoting any Higgs boson. Graphs with the outgoing photon 
attached to the fermions $f,\bar f$ are not shown.}
\label{fi:diags_mumuffa}
\end{figure}
The photonic one-loop and bremsstrahlung diagrams that are relevant 
in $\O(\alpha)$ are collected in 
\reffis{fi:diags_vert}--\ref{fi:diags_mumuffa};
the corresponding counterterm graphs are not shown explicitly.
Of course, virtual and real corrections are separately gauge-invariant,
since they correspond to different final states. However, these
corrections are connected in the domains of soft and collinear photon
emission, and these relations between one-loop diagrams and 
their real counterparts have to be preserved in the introduction
of finite decay widths. Otherwise IR divergences and/or collinear
singularities in the high-energy limit might not cancel properly.
Therefore, we do not separate virtual and
real corrections in the following discussion.

Using the same argument as for the tree-level graphs, the diagrams
of the photonic corrections
can be further divided into gauge-invariant subsets according to the 
boson that is exchanged in the related tree-level graph. Thus, we
have a gauge-invariant subset with internal photons only, one
with Z-boson or $\chi$ exchange, and one for each type of Higgs bosons.
Furthermore, each of these subsets splits into three gauge-invariant
parts according to the charge factors of the two fermions to which
the exchanged or emitted photon is attached. In this way we
distinguish initial-state corrections, final-state corrections, and 
corrections connecting initial and final state.
The relative correction factors are proportional to
$Q_\mu^2$, $Q_f^2$, and $Q_\mu Q_f$, respectively. The corresponding one-loop
diagrams are shown in \reffis{fi:diags_vert}(a), \ref{fi:diags_vert}(b),
and \ref{fi:diags_box}, respectively; the related squared diagrams
of real photon emission should be clear.

The $\O(\alpha_{\mathrm{s}})$ QCD corrections, which are only present
if the fermion $f$ is a quark, are closely related to the photonic 
final-state corrections of ${\cal O}(\alpha)$. The QCD diagrams are
obtained from the photon exchange or emission diagrams upon replacing 
the loop-exchanged or emitted photon by a gluon. The QCD diagrams are,
thus, classified exactly in the same way as the photonic final-state 
corrections. 

\subsection{Treatment of Z-boson and Higgs resonances}

In the following we describe a simple way of introducing 
finite widths for the Z and Higgs bosons in the process \refeq{eq:mumuff}
that retains gauge invariance in the presence of photonic and QCD 
corrections.

As explained in the previous section, each Higgs-boson mass can be
viewed as an independent parameter as long as we inspect only 
photonic and QCD corrections. Thus, a possible way of introducing
finite Higgs-boson widths is provided by substituting the real
Higgs masses squared by the complex pole positions in all Higgs
propagators. Since all relations induced
by gauge invariance, such as the cancellation of the gauge-parameter
dependence in the photon propagator, are of algebraic nature, they
are not destroyed by this substitution. 
For Higgs bosons the difference between barred and unbarred
mass and width parameters will not be experimentally significant,
so that there is no need to distinguish between barred and unbarred 
quantities any further.
Thus, for the Higgs masses appearing in Eq.~\refeq{eq:M0} we set
\beq
\hat M_\varphi^2 = M_\varphi^2-\ri M_\varphi\Gamma_\varphi, 
\qquad \varphi=\Ph,\PH,\PA.
\eeq

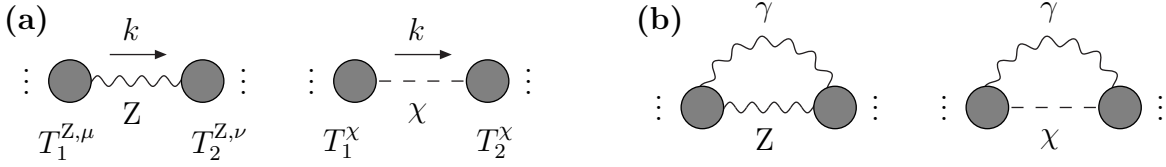
\begin{figure}
\setlength{\unitlength}{1pt}
\centerline{
\begin{picture}(100,40)(0,10)
\Photon   (20,40)(70,40){2}{6}
\GCirc(20,40){8}{.5}
\GCirc(70,40){8}{.5}
\LongArrow(35,50)(55,50)
\Text     ( 40,28)[l]{$\PZ$}
\Text     (  2,43)[l]{\vdots}
\Text     ( 84,43)[l]{\vdots}
\Text     (-5,63)[l]{\bf (a)}
\Text     (  8,17)[l]{$T^{\PZ,\mu}_1$}
\Text     ( 66,17)[l]{$T^{\PZ,\nu}_2$}
\Text     ( 40,60)[l]{$k$}
\end{picture} 
\hspace*{.0em}
\begin{picture}(100,40)(0,10)
\DashLine(20,40)(70,40){5}
\GCirc(20,40){8}{.5}
\GCirc(70,40){8}{.5}
\LongArrow(35,50)(55,50)
\Text     ( 40,28)[l]{$\chi$}
\Text     (  2,43)[l]{\vdots}
\Text     ( 84,43)[l]{\vdots}
\Text     (  8,17)[l]{$T^{\chi}_1$}
\Text     ( 66,17)[l]{$T^{\chi}_2$}
\Text     ( 40,60)[l]{$k$}
\end{picture} 
\hspace*{2.0em}
\begin{picture}(100,65)(0,20)
\Photon   (20,40)(70,40){2}{6}
\PhotonArc(45,40)(25,0,180){2}{8}
\GCirc(20,40){8}{.5}
\GCirc(70,40){8}{.5}
\Text     ( 40,28)[l]{$\PZ$}
\Text     ( 40,76)[l]{$\gamma$}
\Text     (  2,43)[l]{\vdots}
\Text     ( 84,43)[l]{\vdots}
\Text     (-5,73)[l]{\bf (b)}
\end{picture} 
\hspace*{.0em}
\begin{picture}(100,40)(0,20)
\DashLine(20,40)(70,40){5}
\PhotonArc(45,40)(25,0,180){2}{8}
\GCirc(20,40){8}{.5}
\GCirc(70,40){8}{.5}
\Text     ( 40,28)[l]{$\chi$}
\Text     ( 40,76)[l]{$\gamma$}
\Text     (  2,43)[l]{\vdots}
\Text     ( 84,43)[l]{\vdots}
\end{picture} } 
\caption{Subdiagrams with Z and $\chi$ exchange. 
The blobs represent tree-like or loop parts, and the dots stand
for external (on-shell) fermion or photon fields.}
\label{fi:diags_gauge}
\end{figure}

For Z-boson and $\chi$ exchange the situation is much more involved,
since the poles in the longitudinal part of the Z-boson propagator
and the one in the $\chi$ propagator are unphysical and have to
cancel each other properly. We perform this cancellation explicitly
in a general $R_\xi$ gauge, where the free propagators are given by
\beqar
G^{\PZ\PZ}_{\mu\nu}(k) &=& \frac{-\ri}{k^2-\MZ^2+\ri\MZ\GZ(k^2)}
\left(g_{\mu\nu}-\frac{k_\mu k_\nu}{k^2}\right)
-\frac{\ri\xi_\PZ}{k^2-\xi_\PZ\MZ^2}\,\frac{k_\mu k_\nu}{k^2},
\nn\\
G^{\chi\chi}(k) &=& \frac{\ri}{k^2-\xi_\PZ\MZ^2},
\label{eq:Zprop}
\eeqar
with $\xi_\PZ$ denoting the gauge-fixing parameter of the Z field.
Here we have already taken into account that the transverse part of
the Z~propagator receives a contribution of the physical gauge-boson
width after Dyson summation, while the longitudinal part and the
$\chi$ propagator do not. The actual form of the function
$\GZ(k^2)$ for the width is left open for the moment.%
\footnote{Contributions that do not change the resonance structure, such 
as real parts of self-energies or $\gamma Z$ mixing effects, are not
included in the Dyson summation, since they are part of the
purely weak correction not considered in this paper.}
In the diagrams for the tree level, for the one-loop, and for
the bremsstrahlung corrections, shown in 
\reffis{fi:diags_mumuff}--\ref{fi:diags_mumuffa},
the propagators \refeq{eq:Zprop} appear in the two different
ways schematically shown in \reffi{fi:diags_gauge}.
The subdiagrams (a) contain Z or $\chi$ fields on a tree-like line, i.e.\
cutting this line the diagram is decomposed into two disconnected parts;
the subdiagrams (b) contain a Z or $\chi$ line within a loop.
As indicated in \reffi{fi:diags_gauge}, we denote the structures of the blobs
in subdiagrams (a) by $T^{\PZ,\mu}_n$ and $T^\chi_n$ ($n=1,2$).
The functions $T$ are nothing but (photonic parts of) amputated Green functions 
for off-shell Z or $\chi$ fields contracted with the wave functions
for the remaining external fields of $\mu^+$, $\mu^-$, $f$, $\bar f$,
or $\gamma$ which are all on shell. Since subdiagrams (a) form
a subset of gauge-invariant diagrams (see previous section), the
$T$ functions obey the same well-known Ward identities 
(see e.g.\ \citeres{Chanowitz:1985hj,Denner:1996gb}%
\footnote{Note that there is a global sign difference in the Ward identity
in \citere{Denner:1996gb} because of the different sign convention in
the SU(2) gauge coupling $e/\sw$.})
as the full amputated
Green functions that are on-shell contracted for all legs other than
the Z or $\chi$ line:
\beq
-k_\mu T^{\PZ,\mu}_1(-k) = -\ri\MZ T^\chi_1(-k), \qquad
k_\nu T^{\PZ,\nu}_2(k) = -\ri\MZ T^\chi_2(k),
\label{eq:WI1}
\eeq
where $\mp k$ in the argument of $T_{1,2}$ are the incoming momenta of the 
subdiagrams. Using these identities the contribution $\M_{(a)}$ of class (a)
takes the form
\beqar
\M_{(a)} &=& T^{\PZ,\mu}_1(-k) G^{\PZ\PZ}_{\mu\nu}(k) T^{\PZ,\nu}_2(k) + 
T^\chi_1(-k) G^{\chi\chi}(k) T^\chi_2(k)
\nn\\
&=&
\frac{-\ri T^{\PZ,\mu}_1(-k) T^\PZ_{2,\mu}(k)}{k^2-\MZ^2+\ri\MZ\GZ(k^2)} +
\frac{\ri T^{\chi}_1(-k) T^\chi_2(k)
  \left[k^2+\ri\MZ\GZ(k^2)\right]}{k^2\left[k^2-\MZ^2+\ri\MZ\GZ(k^2)\right]},
\eeqar
where the arbitrary gauge parameter $\xi_\PZ$ has dropped out, as
it should be.
For diagrams of class (b) we proceed similarly.
An important difference is, however, that the subdiagrams contained in
the blobs do not obey Ward identities like Eq.~\refeq{eq:WI1} separately,
because the blobs are linked. Instead the whole rest of the diagram
without the Z or $\chi$ line can be viewed as part of an amputated
Green function with external on-shell particles $\mu^+$, $\mu^-$, $f$, 
$\bar f$ and two off-shell Z or $\chi$ legs. These parts of Green
functions, denoted by $T^{\PZ\PZ,\mu\nu}(-k,k)$ and $T^{\chi\chi}(-k,k)$,
obey the Ward identity
\beq
k_\mu k_\nu T^{\PZ\PZ,\mu\nu}(-k,k) = \MZ^2 T^{\chi\chi}(-k,k), 
\label{eq:WI2}
\eeq
as we have also checked explicitly for the box diagrams. Maybe it is
interesting to note that really the sum of all four relevant box diagrams
on both sides of Eq.~\refeq{eq:WI2} is needed for the identity, i.e.\
it is not valid for the pairs of box diagrams obtained by interchanging
Z and $\chi$ lines.
The contribution $\M_{(b)}$ of class (b) can, thus, be written as
\beqar
\M_{(b)} &=& \int\frac{\rd^4k}{(2\pi)^4}\, \left\{
T^{\PZ\PZ,\mu\nu}(-k,k) G^{\PZ\PZ}_{\mu\nu}(k) +
T^{\chi\chi}(-k,k) G^{\chi\chi}(k) \right\}
\nn\\
&=& \int\frac{\rd^4k}{(2\pi)^4}\, \left\{
\frac{-\ri T^{\PZ\PZ,\mu}_{\mu}(-k,k)}{k^2-\MZ^2+\ri\MZ\GZ(k^2)} +
\frac{\ri T^{\chi\chi}(-k,k)
  \left[k^2+\ri\MZ\GZ(k^2)\right]}{k^2\left[k^2-\MZ^2+\ri\MZ\GZ(k^2)\right]}
\right\},
\label{eq:Mb}
\eeqar
where the loop integration has been made explicit.

Now we have to specify how to choose the function $\GZ(k^2)$ describing
the Z-boson width. It is interesting to note that the simple choice
of identifying $\GZ(k^2)$ with the (on-shell) width $\GZ$ is not consistent,
although $\GZ(\MZ^2)=\GZ$ on resonance. The reason for this failure
is an artificial infrared (IR) divergence in the loop integral of $\M_{(b)}$,
which appears in the second term in the curly bracket of Eq.~\refeq{eq:Mb}
for $k^2\to 0$. The divergence is, however, avoided if the 
$k^2$ dependence of $\GZ(k^2)$ is taken into account, since 
$\GZ(k^2\to 0)=0$. We, therefore, set $\GZ(k^2) =\GZ\times k^2/\MZ^2$ 
which is exact for massless decay products and $k^2>0$.
The case $k^2<0$, where $\GZ(k^2)$ actually vanishes, only occurs
in the box loop integrals $\M_{(b)}$, and the effect that we do not set
$\GZ(k^2)$ to zero there is of the order $\alpha\GZ/\MZ$, i.e.\ beyond
our accuracy. The benefit of the chosen form for $\GZ(k^2)$ becomes
obvious after inserting $\GZ(k^2)$ and reparametrizing 
$\M_{(a)}$ and $\M_{(b)}$ by the complex-pole parameters $\bar\MZ$
and $\bar\GZ$:
\beqar
\M_{(a)} &=& \left(1+\ri\gamma_\PZ\right)^{-1} \left\{
\frac{-\ri T^{\PZ,\mu}_1(-k) T^\PZ_{2,\mu}(k)
      +\ri\bar T^{\chi}_1(-k)\bar T^\chi_2(k)}
       {k^2-\bar\MZ^2+\ri\bar\MZ\bar\GZ} \right\},
\nn\\
\M_{(b)} &=& \left(1+\ri\gamma_\PZ\right)^{-1}
\int\frac{\rd^4k}{(2\pi)^4}\, \left\{
\frac{-\ri T^{\PZ\PZ,\mu}_{\mu}(-k,k)
      +\ri\bar T^{\chi\chi}(-k,k)}
       {k^2-\bar\MZ^2+\ri\bar\MZ\bar\GZ} \right\},
\label{eq:Mab}
\eeqar
where we have defined $\gamma_\PZ=\GZ/\MZ$ and 
\beq
\bar T^{\chi}_n(k) = \sqrt{1+\ri\gamma_\PZ} \, T^{\chi}_n(k), \qquad
\bar T^{\chi\chi}(-k,k) = \left(1+\ri\gamma_\PZ\right) \, T^{\chi\chi}(-k,k),
\qquad n=1,2.
\label{eq:bTchi}
\eeq
The rescaling \refeq{eq:bTchi} is equivalent to substituting 
$\MZ^2\to\bar\MZ^2-\ri\bar\MZ\bar\GZ$ in $T^{\chi}_n$ and $T^{\chi\chi}$, 
since these functions are proportional to $\MZ^{-1}$ and $\MZ^{-2}$, 
respectively,
because of the coupling \refeq{eq:cff} of the $\chi$ field to fermions.

In summary, we have shown that the consistent use of the propagators
\refeq{eq:Zprop}, where only the imaginary part, $\MZ\GZ(k^2)$, of 
the transverse Z-boson self-energy is 
Dyson summed, together with
the approximation $\GZ(k^2)=\GZ\times k^2/\MZ^2$ is equivalent
to an overall replacement $\MZ\to\sqrt{\bar\MZ^2-\ri\bar\MZ\bar\GZ}$
in all graphs with Z-boson or $\chi$ exchange before any Dyson
summation is done, up to the global scaling factor 
$\left(1+\ri\gamma_\PZ\right)^{-1}$. 
This observation again shows that no (algebraic) relations induced
by gauge invariance are violated in this procedure.
In our following calculation we make consistent use of the form
\refeq{eq:Mab}, but drop the global scaling factor
$\left(1+\ri\gamma_\PZ\right)^{-1}$, since it influences squared amplitudes 
only at the level $\GZ^2/\MZ^2$ or $\GZ\Gamma_\varphi/(\MZ M_\varphi)$, which
is beyond the accuracy of interest.
This means that we set
\beq
\hat\MZ^2 = \hat M_\chi^2 = \bar\MZ^2-\ri\bar\MZ\bar\GZ
\eeq
in Eqs.~\refeq{eq:cff} and \refeq{eq:MVPS0}.

\subsection{Outlook}

We emphasize that the procedure described in the previous section
cannot be taken over without
modifications to include the genuine weak ${\cal O}(\alpha)$ corrections
(also denoted by ${\cal O}(\GF)$ elsewhere)
where the situation is more involved.
The main complication is due to the fact that weak self-energy,
vertex, and box corrections are all gauge-dependent and
connected by gauge invariance.
These relations have to be preserved by the Dyson summation.
One possibility to introduce finite decay widths in the resonance
propagators without losing gauge invariance is provided by the
{\it pole scheme} \cite{Stuart:1991xk}, where each propagator is 
expanded about its resonance pole and only the part of the residue,
which is gauge-invariant, gets a finite width in the denominator.
Details of the actual application of this scheme can also be found
in \citere{Denner:1996ug}, where it was applied to the Higgs resonance
in $\PZ\PZ\to\PZ\PZ$, including electroweak $\O(\alpha)$ corrections.
In the pole expansions, of course, care has to be taken that the
relations between photonic virtual and real corrections in the
soft and collinear regions are still retained, which might
additionally complicate the application.

The discussion of the previous section raises the question about
a possibility to consistently use complex mass values, which are derived
from the complex locations $\bar M^2-\ri\bar M\bar\Gamma$ of
the propagator poles, as renormalized mass parameters.
If the complex masses for the W and Z~bosons are used in the definition
\refeq{eq:sw} of the weak mixing angle, all generalized Ward identities,
and thus gauge invariance, remains valid. This {\it complex-mass
scheme} has been introduced in \citere{Denner:1999gp} for
lowest-order predictions. However, whether this scheme is also
consistent for the treatment of radiative corrections is not yet
clear, since the use of complex mass and coupling parameters
is field-theoretically questionable.

The field theoretically most desirable solution certainly consists
in a consistent Dyson summation of all propagators (in fixed-order
predictions), since this includes
effects of running couplings directly in the propagator denominators. 
While this procedure in general violates gauge invariance in the
conventionally quantized gauge theory, as explained also above, it was shown 
\cite{Denner:1996gb} that gauge invariance is preserved after
Dyson summation within the background-field formalism \cite{Denner:1994xt}.
However, following this approach in a consistent one-loop
calculation, the finite widths introduced via Dyson-summed one-loop
self-energies are tree-level quantities, i.e.\ the prediction
obtained this way is not yet fully $\O(\alpha)$-corrected.
Although it is not yet clear whether or how this problem can be
solved consistently in a pragmatic way, the background-field
approach seems to be very promising.

The above considerations apply to a full inclusion of all
weak ${\cal O}(\alpha)$ corrections both to resonant and non-resonant
contributions.
If one is only interested in a precise prediction
of the Higgs resonance shapes, a simpler procedure is feasible.
A reasonable approach would be to
include the weak corrections to the Higgs Yukawa couplings
in terms of effective couplings, i.e.\ the coupling parameters
\refeq{eq:higgscoupls} would be dressed by the weak radiative
corrections to the (on-shell) $\varphi f\bar f$ vertices,
and to treat the photonic and QCD corrections as described in this paper.
This is possible for the following reasons.
Firstly, it is known (see also the numerical results below) that the 
interference of the Higgs bosons with the continuous background is totally
negligible, since the former results from the spin-0 and the latter 
from the spin-1 channel. 
Secondly, there is no problem with gauge invariance, because
each set of Higgs-exchange diagrams is gauge-invariant at tree level and
in the presence of photonic and QCD corrections, as explained above.

\section{Photonic radiative corrections}
\label{se:emrcs}

\subsection{Virtual photonic corrections}

The virtual one-loop corrections add coherently to the lowest-order matrix
element $\M_0$ given in Eq.~\refeq{eq:M0}. Denoting the one-loop
contribution to $\M_0$ that is induced by photon exchange by $\M_1^{(\gamma)}$, 
the one-loop-corrected squared matrix element reads
\beq
|\M|^2 = |\M_0|^2 + 2\Re\{\M_1^{(\gamma)} \M_0^*\} + \dots,
\label{eq:M2virt}
\eeq
where the dots stand for the genuine weak corrections of $\O(\GF)$,
not considered in this paper, and higher-order terms.
The one-loop correction $\sigma_1$ to the cross section is obtained by replacing
$|\M_0|^2$ in Eq.~\refeq{eq:sigma0} by the one-loop contribution of
$|\M|^2$. As explained in the previous section, the one-loop correction 
$\M_1^{(\gamma)}$ receives contributions from vertex and box corrections,
\beq
\M_1^{(\gamma)} = \M_{\mathrm{vert}}^{(\gamma)} + \M_{\mathrm{box}}^{(\gamma)}, 
\eeq
which will be considered in the next two subsections in detail.

The actual calculation of the one-loop diagrams has been carried out
using standard techniques. The Feynman graphs have been generated with
{\sl Feyn\-Arts} \cite{Kublbeck:1990xc} and are evaluated in two
independent ways. In the first calculation {\sl Feyn\-Calc}
\cite{Mertig:1991an} is used to express the amplitudes in terms of SME 
and to reduce the occurring tensor integrals
algebraically to scalar integrals with the Passarino--Veltman
algorithm \cite{Passarino:1979jh}. In the second calculation, also
performed with {\sl Mathematica} but without {\sl Feyn\-Calc},
we express the amplitudes in terms of SME and tensor coefficients.
The reduction of the latter to scalar integrals is then done numerically.
The results of the two calculations are in perfect numerical agreement
(i.e.\ within machine accuracy for non-exceptional phase-space points).
In both calculations the scalar integrals are evaluated using the
methods and results of \citeres{'tHooft:1979xw,Beenakker:1990jr,Denner:1993kt},
where ultraviolet divergences are regulated dimensionally
and IR divergences with infinitesimal photon and gluon
masses, $m_\gamma$ and $m_\Pg$, respectively.
The renormalization is carried out in the on-shell renormalization scheme, 
as e.g.\ described in \citere{Denner:1993kt}.

\subsubsection{Vertex corrections}

In the on-shell renormalization scheme
the external self-energy corrections are
exactly cancelled by counterterms. Thus, the virtual vertex corrections
induced by photon exchange are entirely given by the diagrams shown
in \reffi{fi:diags_vert} and the counterterm contribution for the
vertices. Among the relevant renormalization constants,
only the one for the masses, $\delta m_\mu$ and $\delta m_f$, 
and for the fields, $\delta Z^\mu$ and $\delta Z^f$, receive photonic 
corrections, which can be easily derived from the general results given in
\citere{Denner:1993kt}. We immediately give results for the renormalized
vertices.

The vertex functions for the fermionic couplings of a general vector 
boson $\PV$, of a pseudo-scalar boson $\PP$, and of a scalar boson $\PS$
are parametrized by
\beqar
\Gamma^{\PV f}_\rho(q,q') &=&
\ri e\gamma_\rho(v_{\PV f}-a_{\PV f}\gamma_5)
\nn\\[.3em]
&& {}
+ \ri e Q_f^2 \frac{\alpha}{4\pi} \Biggl[ 
\gamma_\rho(v_{\PV f}-a_{\PV f}\gamma_5) \Lambda_1(s,m_f)
+a_{\PV f}\gamma_\rho\gamma_5 \Lambda_2(s,m_f)
\nn\\
&& \qquad {}
+v_{\PV f}\frac{2m_f(q-q')_\rho}{s} \Lambda_3(s,m_f)
+a_{\PV f}\frac{2m_f(q+q')_\rho}{s} \gamma_5 \Lambda_4(s,m_f) \Biggr],
\nn\\[.3em]
\Gamma^{\PP f}(q,q') &=& eg_{\PP f}\gamma_5 
\left[1+Q_f^2\frac{\alpha}{4\pi}\Lambda_5(s,m_f)\right],
\nn\\[.3em]
\Gamma^{\PS f}(q,q') &=& \ri eg_{\PS f}
\left[1+Q_f^2\frac{\alpha}{4\pi}\Lambda_6(s,m_f)\right],
\label{eq:GammaVPS}
\eeqar
where the momenta $q$ and $q'$ of the vertex functions $\Gamma$ 
correspond to the outgoing fermion and antifermion, respectively,
and $s=(q+q')^2$.
Note that we have already assumed that the fermion fields are on shell, 
i.e.\ we have used $q^2=q^{\prime 2}=m_f^2$ and terms that vanish
after contraction with Dirac spinors have been omitted.
The virtual corrections to these vertices are parametrized by the functions 
$\Lambda_i(s,m)$. At the one-loop level these are given by
\beqar
\Lambda_1(s,m) &=& -3\left[B_0(s,m,m)-B_0(0,m,m)-2\right]
       -2(s-2m^2) C_0(m^2,s,m^2,m_\gamma,m,m) 
\nn\\ && {}
+4m^2 B'_0(m^2,m_\gamma,m),
\nn\\[.3em]
\Lambda_2(s,m) &=& 
      \frac{8m^2}{s-4m^2} \left[B_0(s,m,m)-B_0(0,m,m)-2\right],
\nn\\[.3em]
\Lambda_3(s,m) &=& -\frac{s}{s-4m^2} \left[B_0(s,m,m)-B_0(0,m,m)-2\right],
\nn\\[.3em]
\Lambda_4(s,m) &=& -2
       - \frac{3s-4m^2}{s-4m^2}\left[B_0(s,m,m)-B_0(0,m,m)-2\right],
\nn\\[.3em]
\Lambda_5(s,m) &=& \Lambda_1(s,m)-\Lambda_2(s,m)-\Lambda_4(s,m) 
\nn\\
&=& 2 - 2(s-2m^2)C_0(m^2,s,m^2,m_\gamma,m,m) +4m^2 B'_0(m^2,m_\gamma,m),
\nn\\[.3em]
\Lambda_6(s,m) &=& 
2 - \frac{8m^2}{s-4m^2} \left[B_0(s,m,m)-B_0(0,m,m)-2\right]
\nn\\ && {}
- 2(s-2m^2)C_0(m^2,s,m^2,m_\gamma,m,m)
+4m^2 B'_0(m^2,m_\gamma,m).
\label{eq:Lambda}
\eeqar
The appearing scalar integrals are easily calculated to
\beqar
B_0(s,m,m) - B_0(0,m,m) - 2 &=& \beta_s\ln(x_s),
\nn\\[.5em]
C_0(m^2,s,m^2,m_\gamma,m,m) &=&
\frac{1}{\beta_s s}\biggl[ \ln\biggl(\frac{m_\gamma^2}{m^2}\biggr)\ln(x_s)
+\frac{1}{2}\ln^2(x_s)+2\Li(1+x_s)
\nn\\
&& {} \qquad -\frac{\pi^2}{2}-2\pi\ri\ln(1+x_s) \biggr],
\nn\\[.5em]
B'_0(m^2,m_\gamma,m) = 
\frac{\partial B_0}{\partial p^2}(p^2,m_\gamma,m)\bigg|_{p^2=m^2} &=&
-\frac{1}{m^2}\biggl[1+\ln\biggl(\frac{m_\gamma}{m}\biggr)\biggr],
\eeqar
where
\beq
\beta_s = \sqrt{1-\frac{4m^2}{s}+\ri\epsilon}, \qquad
x_s = \frac{\beta_s-1}{\beta_s+1},
\eeq
and $\Li(x)$ denotes the usual dilogarithm.
As indicated in Eq.~\refeq{eq:Lambda}, not all of the functions $\Lambda_i$ 
are independent. Owing to the Ward identity \refeq{eq:WI1},
the correction to the $\chi f\bar f$ vertex, and thus the function $\Lambda_5$,
is completely fixed by the correction to the $\PZ f\bar f$ vertex.

When applied to the process \refeq{eq:mumuff}, the parametrization
\refeq{eq:GammaVPS} can be directly inserted to calculate the final-state 
corrections, while the corrections to the initial-state vertices follow from
obvious substitutions. The complete vertex correction to the amplitude
is given by
\beqar
\M_{\mathrm{vert}}^{(\gamma)} &=& 
\frac{\alpha}{4\pi}\sum_{V=\gamma,\PZ} \Biggl\{
\Big[ Q_\mu^2 \Lambda_1(s,m_\mu)+Q_f^2 \Lambda_1(s,m_f) \Big]\M_{\PV,0} 
\nn\\ && \qquad {}
+ \frac{e^2}{s-\hat\MV^2} \Biggl[
 Q_\mu^2 a_{\PV\mu} \Lambda_2(s,m_\mu) 
  \Big(v_{\PV f}\M^{av}_1-a_{\PV f}\M^{aa}_1\Big)
\nn\\ && \qquad\qquad\qquad\quad {}
+Q_f^2 a_{\PV f} \Lambda_2(s,m_f) 
  \Big(v_{\PV\mu}\M^{va}_1-a_{\PV\mu}\M^{aa}_1\Big)
\nn\\ && \qquad\qquad\qquad\quad {}
+Q_\mu^2 v_{\PV\mu} \Lambda_3(s,m_\mu) \frac{4m_\mu}{s}
  \Big(v_{\PV f}\M^{vv}_4-a_{\PV f}\M^{va}_4+m_f a_{\PV f}\M^{va}_2\Big)
\nn\\ && \qquad\qquad\qquad\quad {}
+Q_f^2 v_{\PV f} \Lambda_3(s,m_f) \frac{4m_f}{s}
  \Big(v_{\PV\mu}\M^{vv}_3-a_{\PV\mu}\M^{av}_3-m_\mu a_{\PV\mu}\M^{av}_2\Big)
\nn\\ && \qquad\qquad\qquad\quad {}
+ a_{\PV\mu}a_{\PV f} \frac{4m_\mu m_f}{s}
  \Big[ Q_\mu^2 \Lambda_4(s,m_\mu)+Q_f^2 \Lambda_4(s,m_f) \Big] \M^{aa}_2
\Biggr] \Biggr\}
\nn\\ && {}
+\frac{\alpha}{4\pi}\sum_{\PP=\chi,\PA} 
\Big[ Q_\mu^2 \Lambda_5(s,m_\mu)+Q_f^2 \Lambda_5(s,m_f) \Big]\M_{\PP,0} 
\nn\\ && {}
+\frac{\alpha}{4\pi}\sum_{\PS=\Ph,\PH}  
\Big[ Q_\mu^2 \Lambda_6(s,m_\mu)+Q_f^2 \Lambda_6(s,m_f) \Big]\M_{\PS,0},
\label{eq:Mvert}
\eeqar
where we made use of the lowest-order structures $\M_{\PV,0}$,
$\M_{\PP,0}$, and $\M_{\PS,0}$ defined in Eq.~\refeq{eq:MVPS0}.
   
\subsubsection{Box corrections}

Compared to the vertex corrections, the calculation of the box diagrams, 
which are shown in \reffi{fi:diags_box}, involves quite a lot of algebra. 
In fact, the full result for the box corrections is too lengthy and
involved to be reported here. 
Instead we give the most important gauge-invariant part of the boxes, 
denoted by $\M_{\mathrm{box,res}}^{(\gamma)}$, which contains the
complete IR structure and all resonant contributions.
To this end, we split the box correction according to
\beq
\M_{\mathrm{box}}^{(\gamma)} = 
\M_{\mathrm{box,res}}^{(\gamma)} + \M_{\mathrm{box,cont}}^{(\gamma)}.
\eeq
The finite remainder $\M_{\mathrm{box,cont}}^{(\gamma)}$,
which describes box corrections to the continuum, will not be
given analytically, but it is included in the numerical results below.

The part $\M_{\mathrm{box,res}}^{(\gamma)}$ is obtained from the full
box diagrams upon setting the integration momentum in the numerator
of the diagram to the value that corresponds to zero-momentum transfer
by the photon. If two photons are exchanged in the box, there are two
such values of the integration momentum, leading to two independent
contributions. Obviously, this procedure extracts the complete
IR-sensitive part of the box diagrams, as can be shown by simple
power counting: if $q_\gamma$ is the photon momentum, the difference
between the full box and the extracted part contains a factor
$q_\gamma$, which cancels the (logarithmic) IR divergence ($q_\gamma\to 0$)
of the denominator. The integration domain near $q_\gamma\to 0$ is, 
however, also the only source of potentially resonant contributions of 
the box diagrams. A resonance requires $s\sim M_X^2$, where $X$ denotes
any boson exchanged at tree level, and the only way to preserve the
condition $q_X^2\sim s\sim M_X^2$ ($q_X$ denoting the momentum of $X$)
in the box diagram is to require
that the photon, which also connects initial and final state, is soft
($q_\gamma\to 0$). Otherwise the $X$ propagator becomes non resonant. 
Since we did not set $q_\gamma\to 0$ in the $X$ propagator denominator in our
definition of $\M_{\mathrm{box,res}}^{(\gamma)}$, the full leading
resonance behaviour of each box is still contained in this part of the
box amplitudes.
The actual calculation of $\M_{\mathrm{box,res}}^{(\gamma)}$ is
rather simple, since it involves only scalar integrals, and the
result is
\beqar
\M_{\mathrm{box,res}}^{(\gamma)} &=&
Q_\mu Q_f \frac{\alpha}{\pi} 
\sum_{X=\gamma,\PZ,\Ph,\PH,\chi,\PA} 
\left[(m_\mu^2+m_f^2-t)D_0(t,\hat M_X) -(t\leftrightarrow u)\right] 
\nn\\
&& \hspace*{9em} {}
\times (s-\hat M_X^2) \M_{X,0} / (1+\delta_{X\gamma}),
\label{eq:Mboxres}
\eeqar
where $\delta_{X\gamma}=1$ for $X=\gamma$ and $\delta_{X\gamma}=0$ otherwise.
The gauge invariance of this part of the one-loop amplitude simply
follows from the gauge invariance of the $Q_\mu Q_f$ part of the 
IR divergence, which is completely contained in 
$\M_{\mathrm{box,res}}^{(\gamma)}$.
The IR-divergent scalar box integrals $D_0$, including the generalization to
complex masses $\hat M_X$, have been calculated in \citere{Beenakker:1990jr};
the results read
\beqar
D_0(t,\hat M_X) &=& 
D_0(m_\mu^2,m_\mu^2,m_f^2,m_f^2,s,t,m_\gamma,m_\mu,\hat M_X,m_f)
\nn\\
&=& \frac{1}{(s-\hat M_X^2)\sqrt{\lambda(t,m_\mu^2,m_f^2)}} \Biggl\{
2\ln(x_t)\left[\ln(1-x_t^2)
    -\ln\left(\frac{\hat M_X m_\gamma}{\hat M_X^2-s}\right) \right]
\nn\\ && \qquad {}
-\sum_{\rho,\sigma=\pm 1} \left[ \Li(x_t x_\mu^\rho x_f^\sigma)
+\Big(\ln(x_t)+\ln(x_\mu^\rho)+\ln(x_f^\sigma)\Big)
    \ln(1-x_t x_\mu^\rho x_f^\sigma) \right]
\nn\\ && \qquad {}
+\frac{\pi^2}{2}+\Li(x_t^2)+\ln^2(x_\mu)+\ln^2(x_f)
\Biggr\} \qquad \mbox{for } X\ne \gamma,
\nn\\[.5em]
D_0(t,\hat M_\gamma) &=& 
D_0(m_\mu^2,m_\mu^2,m_f^2,m_f^2,s,t,m_\gamma,m_\mu,m_\gamma,m_f)
\nn\\
&=& -\frac{1}{s\sqrt{\lambda(t,m_\mu^2,m_f^2)}} \,
2\ln(x_t)\ln\left(-\frac{m_\gamma^2}{s}+\ri\eps\right),
\eeqar
where we have used the usual definition
\beq
\lambda(x,y,z) = x^2+y^2+z^2-2xy-2xz-2yz
\eeq
and the abbreviations
\beqar
x_t &=& \frac{\beta_t-1}{\beta_t+1}, \qquad 
\beta_t=\sqrt{1-\frac{4m_\mu m_f}{t-(m_\mu-m_f)^2}+\ri\eps},
\nn\\
x_a &=& \sqrt{\frac{1-\beta_{aX}}{1+\beta_{aX}}}, \qquad
\beta_{aX} = \sqrt{1-\frac{4m_a^2}{\hat M_X^2}}, \qquad a=\mu,f, 
\quad X\ne\gamma.
\eeqar
Note that the explicit factor $(s-\hat M_X^2)$ in Eq.~\refeq{eq:Mboxres}
is cancelled by the prefactor in the $D_0$ functions, but the
resonance factor in the lowest-order structure $\M_{X,0}$ survives.

It is instructive to inspect the part
$\M_{\mathrm{box,res}}^{(\gamma)}$ from a different point of view.
Each contribution from $X\ne\gamma$ can also be viewed as the leading
term in an expansion of the corresponding box diagram about the
resonance pole. Hence, these terms are nothing but the virtual part
of the so-called {\it non-factorizable} corrections which were discussed
in detail in \citere{Melnikov:1995fx} for several processes with
W and Z~resonances.%
\footnote{Actually $\M_{\mathrm{box,res}}^{(\gamma)}$ differs 
from the non-factorizable corrections by non-resonant contributions,
since the off-shellness $s$ is set to its resonance value in the latter.}
From the results given there
it is also clear that the resonant part of $\M_{\mathrm{box,res}}^{(\gamma)}$
will be completely compensated by its counterpart from soft photon
emission. Thus, we can expect that the sum of virtual and real
initial-final corrections to the resonances will be very small,
as it is also known (see e.g.\ \citere{LEP1} and references therein)
for the Z~resonance in $\Pep\Pem$ collisions for a long time.

Since the continuum contribution $\M_{\mathrm{box,cont}}^{(\gamma)}$ is
gauge-invariant and does not involve resonant terms, there is no
need to introduce complex masses in this part. 
Thus, we keep all masses real in the tensor integrals there. 
Introducing complex masses
would change the cross section by terms of $\O(\alpha\Gamma_X/M_X)$
which is beyond the desired level of accuracy.

If one is only interested in resonance cross sections, a simple
approximation for the corrections in the vicinity of the
resonance at $s\sim M_X^2$ is provided by the neglect of the
continuum contribution $\M_{\mathrm{box,cont}}^{(\gamma)}$.
This approach is known as {\it pole approximation} and, for instance,
successfully applied for Drell--Yan-like W~production at hadron colliders
\cite{Wackeroth:1997hz} and W-pair production in $\Pep\Pem$ annihilation
\cite{Jadach:1998hi}. 
One advantage of this approximation is certainly
its simplicity compared to the full off-shell calculation.

\subsection{Real photonic corrections}

\subsubsection{Amplitudes for $\mu^-\mu^+\to f\bar f\gamma$}
\label{se:mmffa}
\newcommand{\gVfp}{g_{\PV f}^+}
\newcommand{\gVfm}{g_{\PV f}^-}
\newcommand{\gVmup}{g_{\PV\mu}^+}
\newcommand{\gVmum}{g_{\PV\mu}^-}
\newcommand{\gSfp}{g_{\varphi f}^+}
\newcommand{\gSfm}{g_{\varphi f}^-}
\newcommand{\gSmup}{g_{\varphi\mu}^+}
\newcommand{\gSmum}{g_{\varphi\mu}^-}
\newcommand{\KPpPK}     {\langle k P' P k \rangle}
\newcommand{\KQpQK}     {\langle k Q' Q k \rangle}
\newcommand{\MPK}{(p\cdot k)}
\newcommand{\MPpK}{(p'\cdot k)}
\newcommand{\MQK}{(q\cdot k)}
\newcommand{\MQpK}{(q'\cdot k)}
\newcommand{\phiXI} {\langle \phi  \xi'  \rangle}
\newcommand{\phieta}{\langle \phi  \eta  \rangle}
\newcommand{\psixi} {\langle \psi  \xi   \rangle}
\newcommand{\psiETA}{\langle \psi  \eta' \rangle}
\newcommand{\PHIxi} {\langle \phi' \xi   \rangle}
\newcommand{\PHIETA}{\langle \phi' \eta' \rangle}
\newcommand{\PSIXI} {\langle \psi' \xi'  \rangle}
\newcommand{\PSIeta}{\langle \psi' \eta  \rangle}
\newcommand{\XIeta} {\langle \xi'  \eta  \rangle}
\newcommand{\xiETA} {\langle \xi   \eta' \rangle}
\newcommand{\phiPSI}{\langle\phi\psi'\rangle}
\newcommand{\PHIpsi}{\langle\phi'\psi\rangle}
\newcommand{\PHIqphi}{\langle \phi' Q \phi \rangle}
\newcommand{\psiqPSI}{\langle \psi  Q \psi'\rangle}
\newcommand{\ETApeta}{\langle \eta' P \eta \rangle}
\newcommand{\xipXI}  {\langle \xi   P \xi' \rangle}
\newcommand{\CphiPSI}{\phiPSI^*}
\newcommand{\CPHIpsi}{\PHIpsi^*}
\newcommand{\CphiXI} {\phiXI^*}
\newcommand{\Cphieta}{\phieta^*}
\newcommand{\Cpsixi} {\psixi^*}
\newcommand{\CpsiETA}{\psiETA^*}
\newcommand{\CXIeta} {\XIeta^*}
\newcommand{\CxiETA} {\xiETA^*}
\newcommand{\CPHIxi} {\PHIxi^*}
\newcommand{\CPHIETA}{\PHIETA^*}
\newcommand{\CPSIXI} {\PSIXI^*}
\newcommand{\CPSIeta}{\PSIeta^*}
\newcommand{\Kpsi}{\langle k \psi \rangle}
\newcommand{\Kxi} {\langle k \xi \rangle}
\newcommand{\KPHI}{\langle k \phi' \rangle}
\newcommand{\KXI} {\langle k \xi' \rangle}
\newcommand{\KETA}{\langle k \eta' \rangle}
\newcommand{\CKpsi}{\Kpsi^*}
\newcommand{\CKxi} {\Kxi^*}
\newcommand{\CKPHI}{\KPHI^*}
\newcommand{\CKXI} {\KXI^*}
\newcommand{\CKETA}{\KETA^*}

The real photonic ${\cal O}(\alpha)$ corrections are induced by the process
\beq
\mu^-(p,\sigma) + \mu^+(p',\sigma') \;\longrightarrow\;
f(q,\tau) + \bar f(q',\tau') + \gamma(k,\lambda)
\label{eq:mumuffgamma}
\eeq
in lowest order, where $k$ and $\lambda$ denote the momenta and helicity
of the emitted photon, respectively. Typical diagrams for this process
are shown in \reffi{fi:diags_mumuffa}.
For the case of the SM the helicity amplitudes have already been given
in \citere{Dittmaier:1999nn}. In the following we modify and generalize 
these amplitudes for our purposes. For completeness we repeat here the basic
definitions needed to evaluate the amplitudes, but suppress most of the details
concerning their derivation, which was done within the Weyl--van der
Waerden (WvdW) spinor technique. We consistently use the variant
of the WvdW formalism described in \citere{Dittmaier:1999nn}, where 
also references to other spinor techniques are given.

In the WvdW formalism, each of the momenta $p^\mu$, etc., is written
as a complex $2\times2$ matrix $P_{\dot AB}$, etc., consistently
denoted by the respective capital letter in the following.
These matrices are decomposed into Weyl spinors as follows,
\beq
P^{(\prime)}_{\dot AB} = 
  \sum_{i=1,2} \kappa^{(\prime)}_{i,\dot A} \kappa^{(\prime)}_{i,B}, \qquad
Q^{(\prime)}_{\dot AB} = 
  \sum_{i=1,2} \rho^{(\prime)}_{i,\dot A} \rho^{(\prime)}_{i,B}, \qquad
K_{\dot AB} = k_{\dot A} k_{B},
\eeq
where a dotted spinor index indicates complex conjugation. 
Using polar coordinates for the momentum $p$,
\beq
p^\mu = (p_0,|{\bf p}|\cos\phi_p\sin\theta_p,|{\bf p}|\sin\phi_p\sin\theta_p,
         |{\bf p}|\cos\theta_p),
\eeq
the (covariant) spinors $\kappa_{i,A}$ are given by
\beq
\kappa_{1,A} = \sqrt{p_0+|{\bf p}|}
\pmatrix{\mathrm{e}^{-\ri\phi_p}\cos\frac{\theta_p}{2} \cr 
\sin\frac{\theta_p}{2} }
\qquad
\kappa_{2,A} = \sqrt{p_0-|{\bf p}|}
\pmatrix{
\sin\frac{\theta_p}{2} \cr
-\mathrm{e}^{+\ri\phi_p}\cos\frac{\theta_p}{2} }.
\eeq
Moreover, contravariant spinors are defined according to
\beq
\phi^A = \eps^{AB}\phi_B, \qquad
\eps^{AB} = \pmatrix{0&+1\cr -1&0}.
\eeq
The application to $\kappa'_{i,A}$ and $\rho^{(\prime)}_{i,A}$ 
is obvious. The spinor $k_A$ is constructed from the 
momentum $k^\mu$ in the same way as $\kappa_{1,A}$ from $p^\mu$
(the counterpart of $\kappa_{2,A}$ vanishes for light-like momenta).
For two Weyl spinors $\phi_A$ and $\psi_A$ the antisymmetric product
\beq
\langle\phi\psi\rangle = \phi_{A}\psi^{A} =
\phi_1\psi_2-\phi_2\psi_1, \qquad
\langle\phi\psi\rangle^* = \phi_{\dot A}\psi^{\dot A} =
(\phi_1\psi_2-\phi_2\psi_1)^*,
\eeq
turns out to be very useful, since both Minkowski products and
Dirac spinor chains can be written in terms of such products.
For helicity eigenstates of fermions, the Dirac spinors can be
expressed in terms of the associated momentum spinors in a very 
simple way. In our case we can write the Dirac spinors as
\beq
u_{\mu^-} =
\pmatrix{ \vphantom{\phi_{\dot A}}\phi_A \cr \psi^{\dot A} },
\qquad
\bar v_{\mu^+} = (\psi^{\prime A},\phi'_{\dot A}),
\qquad
\bar u_f = (\eta^A,\xi_{\dot A}),
\qquad
v_{\bar f} = 
\pmatrix{ \vphantom{\xi'_{\dot A}}\xi'_A \cr \eta^{\prime\dot A} },
\label{eq:diracsp}
\eeq
with the actual insertions
\newcommand{\xieta}{\langle\xi  \eta\rangle}
\newcommand{\ETAXI}{\langle\eta'\xi'\rangle}
\beqar
(\phi,\psi) &=& \Biggl\{
\barr{ll} (\kappa_1,-\kappa_2) & \quad \mbox{for} \; \sigma=+, \\
          (\kappa_2, \kappa_1) & \quad \mbox{for} \; \sigma=-, \earr
\qquad
(\phi',\psi') = \Biggl\{
\barr{ll} ( \kappa'_1,\kappa'_2) & \quad \mbox{for} \; \sigma'=-, \\
          (-\kappa'_2,\kappa'_1) & \quad \mbox{for} \; \sigma'=+, \earr
\nn\\
(\xi,\eta) &=& \Biggl\{
\barr{ll} (\rho_1,-\rho_2) & \quad \mbox{for} \; \tau=+, \\
          (\rho_2, \rho_1) & \quad \mbox{for} \; \tau=-, \earr
\qquad\;\,
(\xi',\eta') = \Biggl\{
\barr{ll} ( \rho'_1,\rho'_2) & \quad \mbox{for} \; \tau'=-, \\
          (-\rho'_2,\rho'_1) & \quad \mbox{for} \; \tau'=+, \earr
\label{eq:wvdwins}
\eeqar
for the various polarizations. 

The helicity amplitudes $\M_\gamma$ for the process \refeq{eq:mumuffgamma}
read
\beq
\M_\gamma(\sigma,\sigma',\tau,\tau',\lambda) = 
\sqrt{2}e^3 \Biggl[ 
\sum_{\PV=\gamma,\PZ} A^{(\PV)}_\lambda(\sigma,\sigma',\tau,\tau') 
+ \sum_{\varphi=\chi,\Ph,\PH,\PA} 
  A^{(\varphi)}_\lambda(\sigma,\sigma',\tau,\tau') \Biggr],
\label{eq:mumuffaamp}
\eeq
where the two generic functions $A^{(\PV)}_\lambda$ and $A^{(\varphi)}_\lambda$ 
describe spin-1 and spin-0 exchange, respectively.
In the SM the sum over $\varphi$ extends only over $\PH$ and $\chi$.
For $\lambda=+1$ these functions are given by
\beqar
A^{(\PV)}_+(\sigma,\sigma',\tau,\tau') &=&
\left\{ \frac{Q_\mu\KPpPK}{2\MPK\MPpK[(q+q')^2-\hat\MV^2]}
- \frac{Q_f\KQpQK}{2\MQK\MQpK[(p+p')^2-\hat\MV^2]} \right\}
\nn\\ 
&& {} \quad \times \Bigl(
\gVmup\gVfp\CPHIxi\phiXI + \gVmup\gVfm\CPHIETA\phieta 
\nn\\ 
&& \phantom{{} \quad \times \Bigl(}
+ \gVmum\gVfp\Cpsixi\PSIXI + \gVmum\gVfm\CpsiETA\PSIeta \Big)
\nn\\ 
&& {} - \frac{Q_\mu}{(q+q')^2-\hat\MV^2} \biggl[ 
\frac{\gVmup\CKPHI}{\MPpK} \Big( \gVfp\CKxi\phiXI+\gVfm\CKETA\phieta \Big)
\nn\\ 
&& \phantom{{} - \frac{Q_\mu}{(q+q')^2-\hat\MV^2} \biggl[ }
- \frac{\gVmum\CKpsi}{\MPK} \Big( \gVfp\CKxi\PSIXI+\gVfm\CKETA\PSIeta \Big) 
\biggr]
\nn\\ 
&& {} - \frac{Q_f}{(p+p')^2-\hat\MV^2} \biggl[ 
\frac{\gVfp\CKxi}{\MQK} \Big( \gVmup\CKPHI\phiXI+\gVmum\CKpsi\PSIXI \Big)
\nn\\ 
&& \phantom{{} - \frac{Q_f}{(p+p')^2-\hat\MV^2} \biggl[ }
- \frac{\gVfm\CKETA}{\MQpK} \Big( \gVmup\CKPHI\phieta+\gVmum\CKpsi\PSIeta \Big)
\biggr],
\nn\\[.5em]
A^{(\varphi)}_+(\sigma,\sigma',\tau,\tau') &=&
\left\{ \frac{Q_f\KQpQK}{4\MQK\MQpK[(p+p')^2-\hat M_\varphi^2]} 
- \frac{Q_\mu\KPpPK}{4\MPK\MPpK[(q+q')^2-\hat M_\varphi^2]} \right\}
\nn\\ 
&& {} \quad \times \Bigl( \gSmup\phiPSI+\gSmum\CPHIpsi \Big)
\Bigl( \gSfp\XIeta+\gSfm\CxiETA \Big)
\nn\\ 
&& {} + \frac{Q_\mu\gSmum\CKPHI\CKpsi}{(q+q')^2-\hat M_\varphi^2} 
\Bigl( \gSfp\XIeta+\gSfm\CxiETA \Big)
\biggl[ \frac{1}{2\MPK}+\frac{1}{2\MPpK} \biggr]
\nn\\ 
&& {} - \frac{Q_f\gSfm\CKxi\CKETA}{(p+p')^2-\hat M_\varphi^2} 
\Bigl( \gSmup\phiPSI+\gSmum\CPHIpsi \Big)
\biggl[ \frac{1}{2\MQK}+\frac{1}{2\MQpK} \biggr].
\nn\\
\eeqar
Here we made consistent use of the chiral couplings introduced in 
Eq.~\refeq{eq:chiralcoupls}, and the dots stand for ordinary
products of four-vectors. Moreover, the following abbreviations are used,
\beq
\langle k P' P k \rangle = \sum_{i,j=1,2}
\langle k \kappa'_i \rangle^* \langle \kappa_j \kappa'_i \rangle
\langle \kappa_j k \rangle^*,
\qquad
\langle k Q' Q k \rangle = \sum_{i,j=1,2}
\langle k \rho'_i \rangle^* \langle \rho_j \rho'_i \rangle
\langle \rho_j k \rangle^*.
\eeq
The amplitudes for $\lambda=-1$ follow from discrete symmetries.
For instance, a parity transformation leads to the relations
\beq
\M_\gamma(-\sigma,-\sigma',-\tau,-\tau',-\lambda) =
{\mathrm{sgn}}(\sigma\sigma'\tau\tau') 
\M_\gamma(\sigma,\sigma',\tau,\tau',\lambda)^*
\Big|_{ (g_{\dots}^\pm)^* \to g_{\dots}^\mp, 
        \hat M_{\dots}^* \to \hat M_{\dots} },
\label{eq:mumuffaP}
\eeq
and CP implies
\beq
\M_\gamma(-\sigma',-\sigma,-\tau',-\tau,-\lambda) =
-{\mathrm{sgn}}(\sigma\sigma'\tau\tau') 
\M_\gamma(\sigma,\sigma',\tau,\tau',\lambda)^*
\Big|_{p\leftrightarrow p', q\leftrightarrow q', 
       (g_{\dots}^\pm)^* \to g_{\dots}^\pm,
       \hat M_{\dots}^* \to \hat M_{\dots}}.
\label{eq:mumuffaCP}
\eeq
The substitutions for $(g_{\dots}^\pm)^*$ and $M_{\dots}^*$ ensure
(in addition to the interchange of $g_{\dots}^+$ and $g_{\dots}^-$
in the P transformation) that complex couplings and masses
are not subject of taking the complex conjugate of the amplitude.

The bremsstrahlung contribution $\sigma_\gamma$ to the cross section
is obtained from the transition amplitudes $\M_\gamma$ by the
phase-space integration
\beq
\sigma_\gamma = \frac{1}{2s\beta_\mu} \int \rd\Phi_\gamma \,
\sum_{\sigma,\sigma',\tau,\tau',\lambda}
\frac{1}{4}(1+2P_-\sigma)(1+2P_+\sigma') \,
|\M_\gamma(\sigma,\sigma',\tau,\tau',\lambda)|^2,
\label{eq:hbcs}
\eeq
where the phase-space integral is defined by
\beq
\int \rd\Phi_\gamma =
\int\frac{\rd^3 {\bf q}}{(2\pi)^3 2q_{0}}
\int\frac{\rd^3 {\bf q}'}{(2\pi)^3 2q'_{0}}
\int\frac{\rd^3 {\bf k}}{(2\pi)^3 2k_0} \,
(2\pi)^4 \delta(p+p'-q-q'-k).
\label{eq:dGg}
\eeq

\subsubsection{Treatment of infrared singularities}

The phase-space integral \refeq{eq:hbcs} of the real photonic
corrections diverges in the IR region, i.e.\ if the photon energy
$k_0$ tends to zero. As done for the virtual corrections, we
regularize this IR divergence by an infinitesimal photon mass $m_\gamma$,
i.e.\ the integral \refeq{eq:hbcs} has to be modified appropriately.
We have applied two independent methods for this task, known
as {\it phase-space slicing} and {\it subtraction}.

In the phase-space slicing approach, the soft-photon region
$m_\gamma<k_0<\Delta E$ is excluded from the integral \refeq{eq:hbcs}
by the energy cut $\Delta E$. If $\Delta E$ is chosen sufficiently
small the photon-emission cross section can be treated in the
soft-photon approximation, where it factorizes into the differential
lowest-order cross section $\rd\sigma_0$ and a universal eikonal
factor (see, e.g., \citere{Denner:1993kt}). Integration over $k$
in the soft-photon region yields the simple correction factor
$\delta_{\soft}$ to $\rd\sigma_0$,
\beq
\delta_{\soft} = \frac{\alpha}{\pi} \left\{
Q_\mu^2 f(\beta_\mu) + Q_f^2 f(\beta_f)
+Q_\mu Q_f \left[ g(t)-g(u) \right] \right\},
\label{eq:dsoft}
\eeq
where the auxiliary functions $f$ and $g$ read
\beqar
f(\beta) &=&
-\ln\left(\frac{4\Delta E^2}{m_\gamma^2}\right)
-\frac{1}{\beta}\ln\left(\frac{1-\beta}{1+\beta}\right)
+\frac{1+\beta^2}{2\beta}\Biggl[
- \ln\left(\frac{4\Delta E^2}{m_\gamma^2}\right) 
  \ln\left(\frac{1-\beta}{1+\beta}\right)
\nn\\ && \qquad {}
+ 2\Li\left(\frac{1-\beta}{1+\beta}\right)
- \frac{1}{2}\ln^2\left(\frac{1-\beta}{1+\beta}\right) - \frac{\pi^2}{3}
+ 2\ln\left(\frac{1-\beta}{1+\beta}\right) 
   \ln\left(\frac{2\beta}{1+\beta}\right)   \Biggr],
\nn\\[.5em]
g(t) &=& 
\frac{2(m_\mu^2+m_f^2-t)}{\sqrt{\lambda(t,m_\mu^2,m_f^2)}} \Biggl[
\ln\left(\frac{a_t m_\mu}{m_f}\right) 
\ln\left(\frac{4\Delta E^2}{m_\gamma^2}\right)
+\frac{1}{4}\ln^2\left(\frac{1-\beta_\mu}{1+\beta_\mu}\right)
-\frac{1}{4}\ln^2\left(\frac{1-\beta_f}{1+\beta_f}\right)
\nn\\ && \qquad {}
+\Li\left(1-\frac{1+\beta_\mu}{v_t}a_t\right)
+\Li\left(1-\frac{1-\beta_\mu}{v_t}a_t\right)
-\Li\left(1-\frac{1+\beta_f}{v_t}\right)
\nn\\ && \qquad {}
-\Li\left(1-\frac{1-\beta_f}{v_t}\right)
\Biggr],
\eeqar
with the abbreviations
\beq
a_t = \frac{m_\mu^2+m_f^2-t+\sqrt{\lambda(t,m_\mu^2,m_f^2)}}{2m_\mu^2},
\qquad
v_t = \frac{2(a_t^2 m_\mu^2-m_f^2)}{s(a_t-1)},
\eeq
and the analogous quantities $a_u$ and $v_u$, obtained from
$a_t$ and $v_t$ by $t\to u$. It can be checked easily that all
IR-singular logarithms $\ln m_\gamma$ of $\delta_{\soft}$ are
compensated by their counterparts of the virtual corrections.
When adding the parts of $\sigma_\gamma$ obtained from the soft
and hard regions, the logarithms of the auxiliary cut $\Delta E$
cancel in the limit $\Delta E\to 0$. Note that in the hard regime
these logarithms result from a numerical integration.
The cancellation proceeds
up to terms of $\O(\Delta E/\varepsilon)$, where $\varepsilon$
is a typical scale in the process. Since $\varepsilon\sim\Gamma_\varphi$
is of the order of some MeV in the vicinity of the light Higgs 
resonances, the cut $\Delta E$ has to be chosen extremely small
there in order to achieve reasonable accuracy in the phase-space
integration of $\sigma_\gamma$.

This numerical subtlety is circumvented in subtraction
methods. Here, the general idea is to subtract and
to add a simple auxiliary function from the singular integrand.
This auxiliary function has to be chosen such that it cancels all
singularities of the original integrand so that the phase-space
integration of the difference can be performed numerically.
Moreover, the auxiliary function has to be simple enough so that it can
be integrated over the singular regions analytically, when the
subtracted contribution is added again.

We have applied the subtraction method presented in
\citere{Dittmaier:2000mb}, where the so-called {\it dipole formalism},
originally introduced by Catani and Seymour \cite{Catani:1996jh}
within massless QCD, was applied to photon radiation and generalized
to massive fermions. Since the actual implementation of this
formalism in our calculation does not involve any subtleties, we do 
not describe it here but refer to \citere{Dittmaier:2000mb} for
details. In particular, the application to the process
$\mu^-\mu^+\to\nu_\Pe\bar\nu_\Pe$, which receives the
same initial-state corrections as the general $f \bar f$ production
process \refeq{eq:mumuff}, is spelled out there explicitly.

The numerical results shown below have been obtained using the
subtraction approach, since the corresponding Monte Carlo integration errors 
are considerably smaller than for the slicing method if the same number
of integration points is used in either case. We have checked that
the results of the two methods agree within the integration errors.

\subsubsection{Leading initial-state radiation beyond $\O(\alpha)$}
\label{se:isr}

The emission of photons collinear to incoming high-energetic charged
fermions, such as $\Pe^\pm$ or $\mu^\pm$ with energies of 
at least several
$10\GeV$, leads to corrections that are enhanced by large logarithms.
These leading logarithms are process independent, and their contribution
to the cross section can be obtained by a convolution of the lowest-order
cross section with a structure function, which is explicitly known
up to $\O(\alpha^3)$ \cite{sf,Beenakker:1996kt}. Following the conventions of
\citere{Beenakker:1996kt}, this convolution reads
\beq
\sigma_{\LL}(s) = \int^1_0\rd x\,\phi(2\alpha,x,Q^2)\,\sigma_0(x s),
\label{eq:sigmaLL}
\eeq
where $\sigma_0(x s)$ is the lowest-order cross section at the reduced
CM energy squared $xs$. The flux function $\phi(2\alpha,x,Q^2)$ 
accounts for radiation from both initial-state muons and is connected
to the QED splitting function $\Gamma^{\LL}$, which 
is up to order $\O(\alpha^3)$ given by
\newcommand{\zetal}{\be_{\mathrm{ISR}}}
\beqar
  \Gamma^{\LL}(x,Q^2) &=& \phi(\alpha,x,Q^2)
\nn\\
&=& \frac{\exp\left(-\frac{1}{2}\zetal\gamma_{\rE} +
        \frac{3}{8}\zetal\right)}
{\Gamma\left(1+\frac{1}{2}\zetal\right)}
    \frac{\zetal}{2} (1-x)^{\frac{\zetal}{2}-1} - \frac{\zetal}{4}(1+x)
\nn\\
&&  {} - \frac{\zetal^2}{32} \biggl\{ \frac{1+3x^2}{1-x}\ln(x)
    + 4(1+x)\ln(1-x) + 5 + x \biggr\}
\nn\\
&&  {} - \frac{\zetal^3}{384}\biggl\{
      (1+x)\left[6\Li(x)+12\ln^2(1-x)-3\pi^2\right]
\nn\\
&& \quad\quad {}
+\frac{1}{1-x}\biggl[ \frac{3}{2}(1+8x+3x^2)\ln(x)
+6(x+5)(1-x)\ln(1-x)
\nn\\
&& \quad\quad\quad {}
+12(1+x^2)\ln(x)\ln(1-x)-\frac{1}{2}(1+7x^2)\ln^2(x)
\nn\\
&& \quad\quad\quad  {}
+\frac{1}{4}(39-24x-15x^2)\biggr] \biggr\},
\label{eq:SFexp}
\eeqar
where $\zetal$ contains the leading logarithm (LL),
\beq
\zetal = Q_\mu^2\frac{2\alpha}{\pi} \left(\ln\frac{Q^2}{m_\mu^2}-1\right).
\eeq
Note that the scale $Q^2$ is not fixed within LL approximation, but
has to be set to a typical scale of the underlying process; for the
numerics we use $Q^2=s$. 
In Eq.~\refeq{eq:SFexp} $\gamma_{\mathrm{E}}$ is the 
Euler constant and $\Gamma(y)$ the gamma function, which should not be
confused with the splitting functions. Note that some non-leading
terms are incorporated, taking into account the fact that the residue
of the soft-photon pole is proportional to $\ln(\dots)-1$ rather than 
the logarithm for the initial-state photon radiation.

We add the cross section \refeq{eq:sigmaLL} to the one-loop result and
subtract the lowest-order and one-loop contributions
$\sigma_{\LL,1}$ already contained
within this formula,
\beqar
\sigma_{\LL,1} &=& \int^1_0 \rd x\,
  \left[\de(1-x)+2\Gamma^{\LL,1}(x,Q^2)\right] \sigma_0(xs),
\eeqar
in order to avoid double counting.
The one-loop contribution to the structure function reads
\beqar
  \Gamma^{\LL,1}(x,Q^2) &=&
  \frac{\zetal}{4} \left(\frac{1+x^2}{1-x}\right)_+
\nl
  &=& \frac{\zetal}{4} \lim_{\eps\to 0}
  \left[\delta(1-x)\left(\frac{3}{2}+2\ln\eps\right)
  + \theta(1-x-\eps)\frac{1+x^2}{1-x}\right].
\label{eq:sigmaLL1}
\eeqar
In summary, the radiatively corrected cross section $\sigma$ reads
\beq
\sigma = \sigma_0 + \sigma_1^{(\gamma)} + \sigma_\gamma
+ \left(\sigma_{\LL}-\sigma_{\LL,1}\right),
\eeq
where $\sigma_1^{(\gamma)}$ is the virtual photonic one-loop
correction.
Note that the uncertainty that is connected with the choice of $Q^2$
enters now in ${\cal O}(\alpha^2)$, since all photonic $\Oa$ corrections,
including constant terms, are taken into account.

\section{QCD radiative corrections}
\label{se:qcdrcs}

As already explained in \refse{se:RCsurvey}, 
the diagrams for the $\O(\alpha_{\mathrm{s}})$ QCD corrections are
obtained from the diagrams for the photonic final-state corrections
upon replacing the loop-exchanged or emitted photon by a gluon,
provided $f$ is a quark.
Simple colour algebra shows that the photonic final-state
corrections turn into the QCD corrections upon replacing the 
coupling factor $Q_f^2\alpha$ by $\CF\alpha_{\mathrm{s}}$, where 
$\CF=4/3$ is the quadratic Casimir operator of the quark 
representation. More precisely, this substitution applies to
the matrix element $\M_{\mathrm{vert}}^{(\gamma)}$ for the
vertex correction in Eq.~\refeq{eq:Mvert}, to the squared matrix 
element $|\M_\gamma|^2$ for hard-photon emission obtained from 
Eq.~\refeq{eq:mumuffaamp}, and 
to the soft-photon factor $\delta_{\soft}$ of Eq.~\refeq{eq:dsoft}.
The contributions proportional to $Q_\mu^2$ or $Q_\mu Q_f$, of course,
have to be set to zero there.
We always evaluate the running strong coupling constant 
$\alpha_{\mathrm{s}}(\mu)$ at the two-loop level 
including five active flavours in the running, i.e.\ the top quark is
decoupled from the running of $\alpha_{\mathrm{s}}(\mu)$.
The renormalization scale is set to $\mu=\sqrt{s}$, which is the total
CM energy. 

The case that $f$ is a b~quark is of particular importance
phenomenologically. For this case it is known that the Higgs
Yukawa couplings $\varphi\Pb\bar\Pb$ receive very large QCD corrections
of the form $\alpha_{\mathrm{s}}\ln(\Mb/\MH)$
that are connected with the on-shell mass renormalization of the
b~quark%
\footnote{Actually, analogous logarithms are present in the photonic
FSR corrections, but there is no need for a special treatment of
those corrections, since they scale with the much smaller factor
$Q_f^2\alpha$ instead of $\CF\alpha_{\mathrm{s}}$.}.
The convergence of the perturbative series is improved
considerably after introducing the $\overline{\mathrm{MS}}$ running b~mass
$\bar\Mb(\mu)$ at an appropriate scale $\mu$. To this end, we
rescale the $\varphi\Pb\bar\Pb$ coupling $g_{\varphi\Pb}$ in all
matrix elements according to
\beq
g_{\varphi\Pb} \;\to\; \bar g_{\varphi\Pb}(s) = 
g_{\varphi\Pb} \times \frac{\bar\Mb\left(\sqrt{s}\right)}{\Mb},
\qquad \varphi=\Ph,\PH,\PA,
\label{eq:Yukrun}
\eeq
which replaces the on-shell mass $\Mb$ by the running b~mass
$\bar\Mb\left(\sqrt{s}\right)$ renormalized at the scale $\mu=\sqrt{s}$.
Note that $\mu^2=s$ corresponds to the virtuality of the
$s$-channel Higgs boson $\varphi$.
The running b~mass is evaluated following the results of
\citere{Chetyrkin:1997dh} (which are also used in {\sl HDECAY}).
In order to avoid double counting, the $\O(\alpha_{\mathrm{s}})$
contribution of the replacement \refeq{eq:Yukrun}
has to be subtracted from the one-loop matrix element $\M_1$. Using 
\beq
\bar\Mb(\mu) = \Mb\left\{1-\frac{\alpha_{\mathrm{s}}(\mu)}{\pi}
\left[\ln\left(\frac{\mu^2}{\Mb^2}\right)+\frac{4}{3}\right]
+\dots \right\},
\eeq
the consistent modification reads
\beq
\M_1 \;\to\; \M_1\Big|_{g_{\varphi\Pb}\to\bar g_{\varphi\Pb}(s)}
+\frac{\alpha_{\mathrm{s}}\left(\sqrt{s}\right)}{\pi}
\left[\ln\left(\frac{s}{\Mb^2}\right)+\frac{4}{3}\right]
\sum_{\varphi=\Ph,\PH,\PA} \M_{\varphi,0}
\Big|_{g_{\varphi\Pb}\to\bar g_{\varphi\Pb}(s)}.
\eeq

Finally, we remark that we have compared our results
against the known analytic form of the inclusive next-to-leading
order QCD corrections \cite{Braaten:1980yq} to the on-shell decays
of scalar and pseudo-scalar Higgs bosons into heavy quarks.
Switching off all non-resonant diagrams in the calculation of the
QCD corrections to the resonance cross sections, the relative QCD
corrections to the peak cross sections reduce to the corrections to
the corresponding partial Higgs decay widths.
We find agreement between our numerical results for the relative
corrections to the peak cross sections and the corresponding analytical
results for the partial Higgs decays within the Monte Carlo integration errors.

\section{Numerical results}
\label{se:num}

\subsection{Input parameters}
\label{se:input}

For the numerical evaluation we use the following set of SM
parameters \cite{Groom:in},
\beq
\begin{array}[b]{lcllcllcl}
\GF & = & 1.16639 \times 10^{-5} \GeV^{-2}, \hspace*{1em}&
\alpha(0) &=& 1/137.0359895, \\
\alpha_{\mathrm{s}}(\MZ) &=& 0.1181, \\
\MW & = & 80.419\GeV, &
\GW & = & 2.12\GeV, \\
\MZ & = & 91.1882\GeV, &
\GZ & = & 2.4952\GeV, \\
m_{\mu} & = & 105.658357\MeV, &
m_{\tau} & = & 1.77703\;\GeV, \\
\Mb & = & 4.62\;\GeV, &
\Mt & = & 174.3\;\GeV, \\
\bar\Ms(1\GeV) & = & 190\;\MeV, &
\Mc & = & 1.42\;\GeV. 
\end{array}
\label{eq:SMpar}
\eeq
Note that part of the input \refeq{eq:SMpar} is only needed for
the calculation of Higgs-boson parameters for which 
{\sl HDECAY} \cite{Djouadi:1997yw} is used; in particular, this
is the case for $\bar\Ms(1\GeV)$ which is the 
$\overline{\mathrm{MS}}$ running strange mass at the renormalization
scale $1\GeV$.

In order to absorb part of the renormalization effects in
the electroweak couplings, such as the running of $\alpha(Q^2)$ 
from $Q^2=0$ to a high-energy scale and some universal effects of the
$\rho$ parameter, we derive the electromagnetic coupling 
$\alpha=e^2/(4\pi)$ from the Fermi constant $\GF$ according to
\beq
\alpha_{\GF} = \frac{\sqrt{2}\GF\MW^2\sw^2}{\pi},
\eeq
when calculating the couplings for the lowest-order cross sections.
In the relative photonic corrections we use $\alpha(0)$ as
coupling parameter, which is the correct effective coupling for
real photon emission. 
Thus, the $\O(\alpha)$-corrected
cross section scales like $\alpha(0)\alpha_{\GF}^2$.
In the following both lowest-order and corrected cross sections 
for $\mu^-\mu^+\to\Pb\bar\Pb$
are consistently evaluated with a running b-quark mass, as described in 
\refse{se:qcdrcs}.

In the MSSM, we take $\tan\beta$ and the mass $\MA$ as the basic
input parameters of the Higgs sector. For the evaluation of the
Higgs decay widths we need additional input parameters, which are
chosen as 
\beqar
\mu &=& 300\GeV, \qquad 
M_2 = 200\GeV, 
\nn\\[.3em]
M_{\tilde l_L} &=& 
M_{\tilde e_R} = 
M_{\tilde q_L} = 
M_{\tilde u_R} = 
M_{\tilde d_R} = 
M_{\tilde L_L} = 
M_{\tilde \tau_R} = 
M_{\tilde Q_L} = 
M_{\tilde t_R} = 
M_{\tilde b_R} = 1\TeV,
\nn\\[.3em]
A_\tau &=& A_t = A_b = 1.5\TeV,
\eeqar
where the notation follows the description of {\sl HDECAY} 
\cite{Djouadi:1997yw}.

Table~\ref{tab:GH} contains the results, as obtained from {\sl HDECAY},
for the Higgs-boson masses and decay widths, which are used in 
the following SM and MSSM evaluations.
\begin{table}
\centerline{
\begin{tabular}{|c|c||c|c|}
\hline
SM & $\MH[\GeV]$ & 115 & 150 
\\ \cline{2-4}
& $\Gamma_\PH[\GeV]$ & 0.003227 & 0.01687 
\\ \hline \hline 
MSSM & $\tan\beta$ & 30 & 5
\\ \cline{2-4}
& $\MA[\GeV]$        & 140     & 400 
\\ \cline{2-4}
& $\Gamma_\PA[\GeV]$ & 2.749   & 0.7939  
\\ \cline{2-4}
& $\Mh[\GeV]$        & 120.995 & 115.948 
\\ \cline{2-4}
& $\Gamma_\Ph[\GeV]$ & 0.1047  & 0.004083
\\ \cline{2-4}
& $\MH[\GeV]$        & 140.658 & 401.975 
\\ \cline{2-4}
& $\Gamma_\PH[\GeV]$ & 2.640   & 0.4726  
\\ \hline
\end{tabular} }
\caption{Higgs-boson masses and decay widths as provided by {\sl HDECAY}}
\label{tab:GH}
\end{table}

\subsection{\boldmath{The process $\mu^-\mu^+\to\Pb\bar\Pb$
in the SM}}

We start our numerical discussion by considering the polarized
cross sections for $\mu^-\mu^+\to\Pb\bar\Pb$ in the SM for $\MH=115\GeV$.
The l.h.s.\ of 
\reffi{fig:mmbb_global_pol} shows the lowest-order cross sections and
their (photonic and QCD) corrected counterparts. 
We first concentrate on the lowest-order cross sections,
where the Higgs-boson resonance occurs only
in the spin-0 channel, i.e.\ for $\mu^\pm$ with equal helicities,
where it appears as a sharp peak at $s=\MH^2$.
Apart from the Higgs resonance region, spin-0 contributions to
the squared lowest-order matrix element $|\M_0|^2$ are suppressed
by a factor $m_\mu^2/\MZ^2\sim 10^{-6}$ with respect to the spin-1
contributions ($\mu^\pm$ with opposite helicities). This suppression factor
is clearly visible in the polarized cross sections. At the Higgs
resonance an enhancement factor $\MH^2/\Gamma_\PH^2$ in the squared
amplitude $|\M_{\PH,0}|^2$
compensates this suppression, as long as the total Higgs width
does not become too large, which is the case if the Higgs decay into
weak gauge bosons is not yet open. In the latter case ($\MH\gsim 2\MW$),
no Higgs resonance will be visible anymore over the continuous
background induced by the spin-1 contribution for realistic degrees
of muon beam polarization.
Therefore, for the SM we consider $\MH=115\GeV$ and $150\GeV$ below.
\bfi
\setlength{\unitlength}{1cm}
\centerline{
\begin{picture}(15.5,8.0)
\put(-5.0,-15.1){\includegraphics{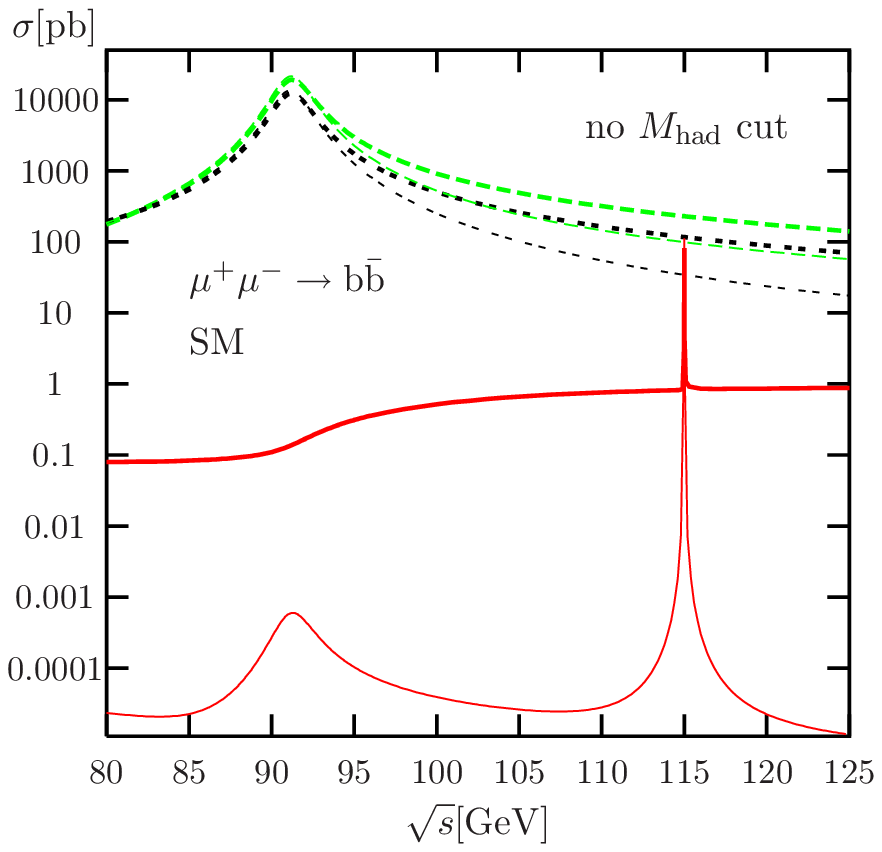}}
\put( 3.0,-15.1){\includegraphics{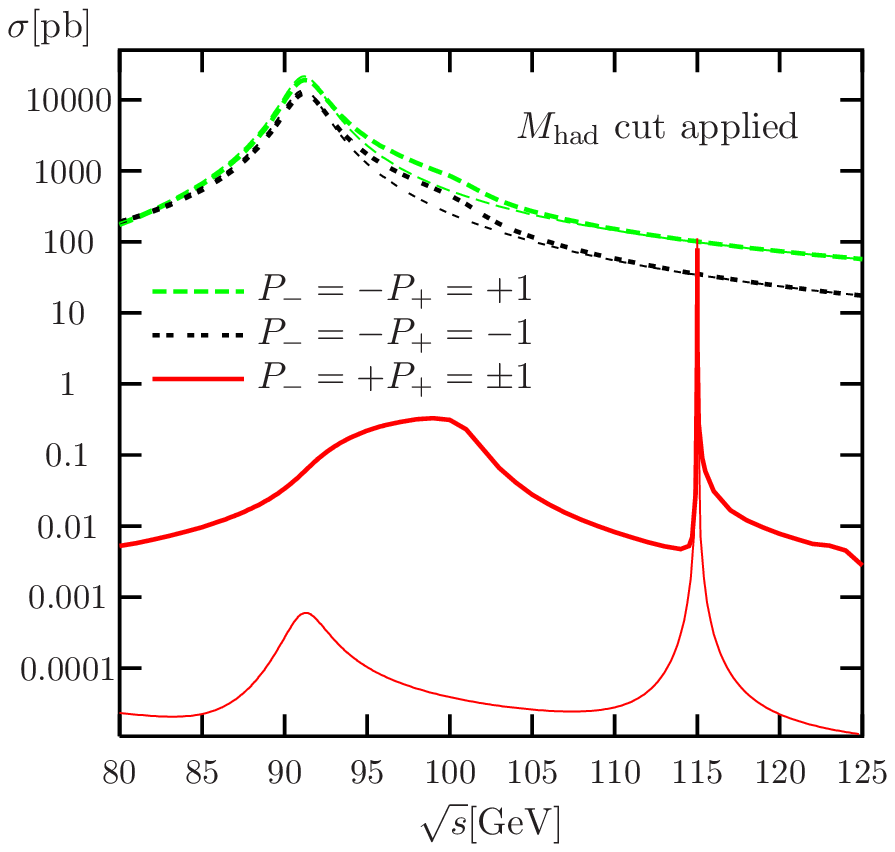}}
\end{picture} } 
\caption{Polarized SM cross sections for $\mu^-\mu^+\to\Pb\bar\Pb$
in lowest order (thin curves) and including photonic and QCD corrections
(thick curves), where no phase-space cuts are applied on the l.h.s.\ and
the invariant-mass cut \refeq{eq:Mhad} is applied on the r.h.s.}
\label{fig:mmbb_global_pol}
\vspace*{3em}
\centerline{
\begin{picture}(8.0,8.0)
\put(-5.0,-15.1){\includegraphics{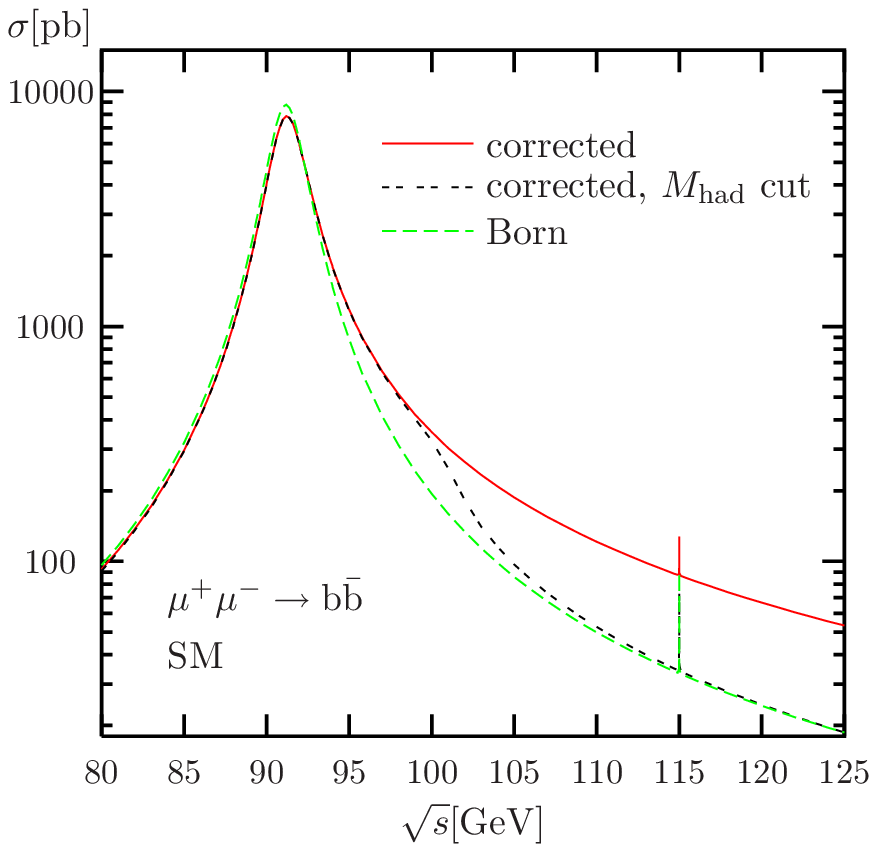}}
\end{picture} } 
\caption{Unpolarized SM cross section for $\mu^-\mu^+\to\Pb\bar\Pb$
in lowest order and including photonic and QCD corrections,
with and without the invariant-mass cut \refeq{eq:Mhad}}
\label{fig:mmbb_global_unpol}
\efi

\begin{sloppypar}
Now we turn to the photonic and QCD corrections to the polarized
cross sections, as shown on the l.h.s.\ of \reffi{fig:mmbb_global_pol}.
In the spin-1 channels, the cross sections
are very similar to its counterpart in $\Pep\Pem$ annihilation
\cite{LEP1}, measured at LEP. Above the Z-boson resonance, ISR
induces very large positive corrections by the so-called {\it radiative
return} to the Z~resonance; these corrections even exceed 100\% in
relative size. The main difference between muon and $\Pep\Pem$
collisions is due to the different LL factors
$\beta_{\mathrm{ISR}}$,
which are about 0.06 and 0.11 for $Q^2=\MZ^2$, respectively, i.e.\ the ISR
effects in muon collisions are roughly half as big as in the $\Pep\Pem$ case.
In the spin-0 channels, which contain the Higgs resonance, even much
larger positive corrections are visible. These corrections are mainly due
to a spin flip of the initial-state muons induced by collinear ISR.
Note that this effect is not accounted for by the LL structure-function
approach, which is widely used as an approximation for ISR, since this
spin-flip correction is not enhanced by a large logarithm. The numerical
enhancement is rather due to the extremely large difference between the
spin-0 and spin-1 cross sections. Neglecting (irrelevant) corrections
that are suppressed by mass factors $m_\mu^2/s$, the spin-flip 
contribution to the real correction $\sigma_\gamma$
takes the universal form \cite{Bohm:1993qx}
\beq
\sigma_{\gamma,\mathrm{spin\, flip}}(P_+,P_-,s) =
\frac{\alpha}{2\pi} \int_0^1\rd x\, (1-x)
\Big[ \sigma_0(-P_+,P_-,xs) + \sigma_0(P_+,-P_-,xs) \Big],
\label{eq:spinflipcs}
\eeq
which can be deduced from the factorization of collinear photon emission
(see e.g.\ \citeres{Dittmaier:2000mb,Kleiss:1986ct}).
Thus, the spin-flip correction 
$\sigma_{\gamma,\mathrm{spin\, flip}}(\pm1,\pm1,s)$ to the spin-0 channel
receives large contributions from the lowest-order spin-1 cross sections
$\sigma_0(\pm1,\mp1,xs)$, multiplied by $\alpha/\pi$ and some numerical
factor resulting from the convolution over $x$. 
Although these corrections look dramatic for completely
polarized beams, they are simply part of the large continuous 
background to the Higgs resonance that is induced by the spin-1
channel for realistic degrees of beam polarization.%
\footnote{The relative weight factor between the spin-1 background
and the spin-0 signal cross sections is roughly given by
$(1-P_+ P_-)/(1+P_+ P_-)$, which will be significantly larger than 
$\alpha/\pi$ in practice.}
\end{sloppypar}

The large ISR corrections from the radiative return can be
suppressed by requiring a minimum invariant mass for the
$\Pb\bar\Pb$ pair that is sufficiently higher than the Z-boson mass.
Therefore, we impose the constraint
\beq
\sqrt{s}-M_{\mathrm{had}} < 10\GeV
\label{eq:Mhad}
\eeq
in the following, where $M_{\mathrm{had}}$ is the invariant mass of
all hadrons in the final state, i.e.\ for gluon emission the gluon
momentum is fully included in $M_{\mathrm{had}}$. This cut implies
that the energy loss by ISR cannot become larger than $10\GeV$.
The r.h.s.\ of \reffi{fig:mmbb_global_pol} shows the lowest-order
and corrected polarized cross sections in the presence of this
cut. The corrections to the spin-1 channel reduce to moderate size,
as expected, but in the spin-0 channels still large corrections 
remain owing to the spin-flip effect described above.

The effect of the photonic and QCD corrections
on the unpolarized cross section for $\mu^-\mu^+\to\Pb\bar\Pb$ is
illustrated in \reffi{fig:mmbb_global_unpol} for $\MH=115\GeV$.
Apart from the Higgs resonance region, the unpolarized cross section
reflects the behaviour of the spin-1 channel discussed above in detail.
In the following we restrict our discussion to the case of unpolarized 
muon beams ($P_\pm=0$), and we always apply the invariant-mass cut 
\refeq{eq:Mhad} for $\Pb\bar\Pb$ production.

Figures~\ref{fig:mmbb_sm_mh115} and \ref{fig:mmbb_sm_mh150}
show close-ups of the Higgs resonances for $\MH=115\GeV$ and
$\MH=150\GeV$, respectively.
\bfi
\setlength{\unitlength}{1cm}
\centerline{
\begin{picture}(15.5,7.7)
\put(-5.0,-15.1){\includegraphics{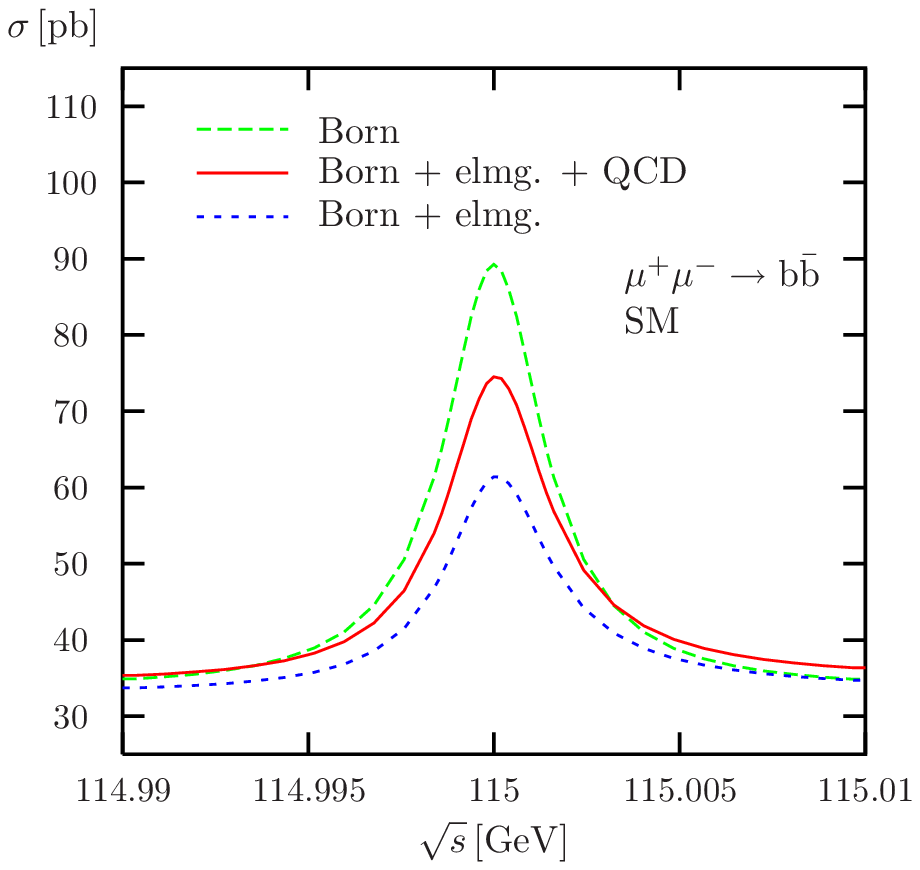}}
\put( 3.0,-15.1){\includegraphics{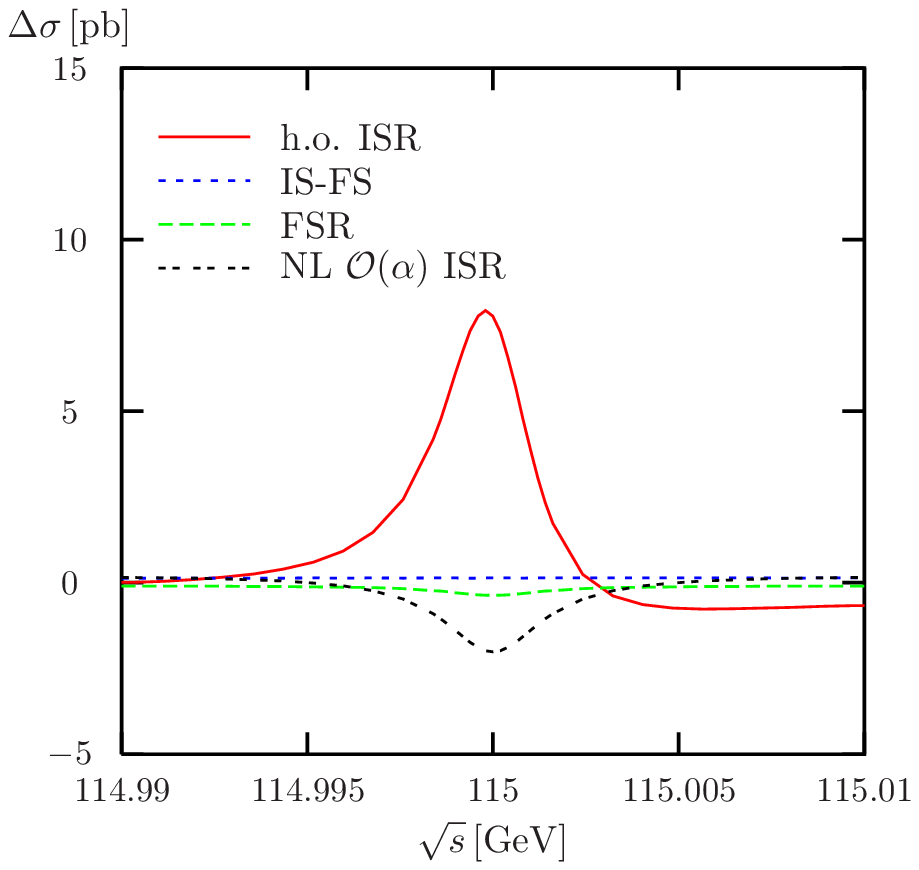}}
\end{picture} } 
\caption{SM cross section $\mu^-\mu^+\to\Pb\bar\Pb$ near the Higgs
resonance for $\MH=115\GeV$ (left) and some contributions
to the photonic corrections (right):
higher-order ISR (h.o.\ ISR), initial-final interferences (IS-FS),
FSR, and non-leading ${\cal O}(\alpha)$ ISR corrections 
(NL ${\cal O}(\alpha)$ ISR)}
\label{fig:mmbb_sm_mh115}
\vspace*{3em}
\centerline{
\begin{picture}(15.5,7.7)
\put(-5.0,-15.1){\includegraphics{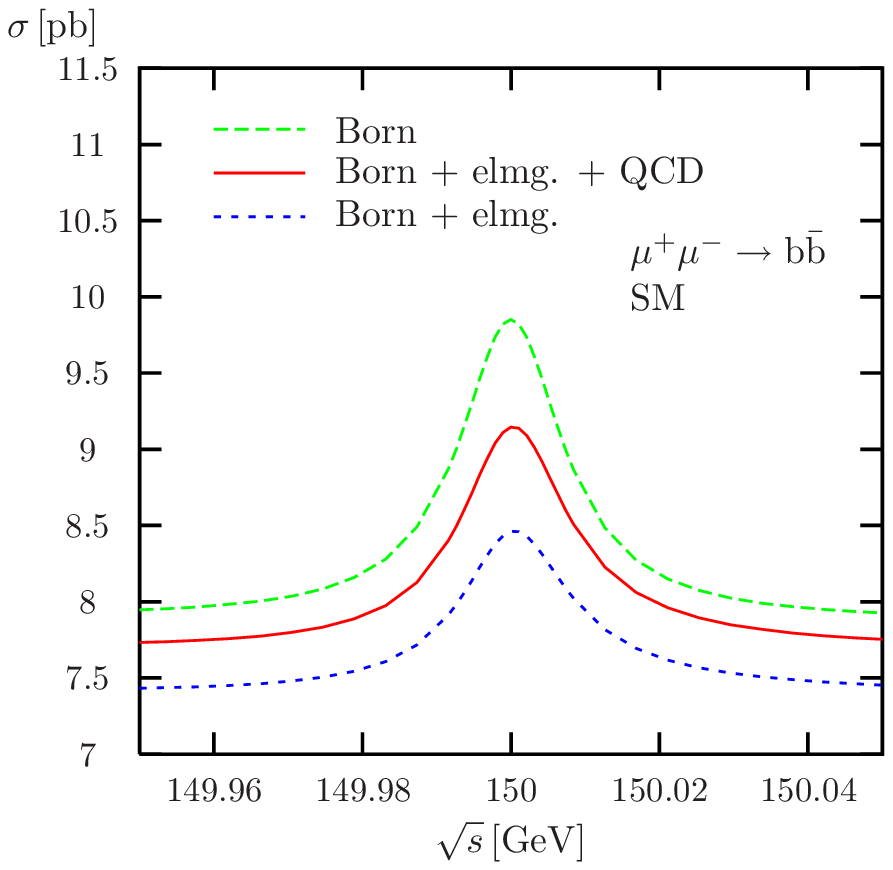}}
\put( 3.0,-15.1){\includegraphics{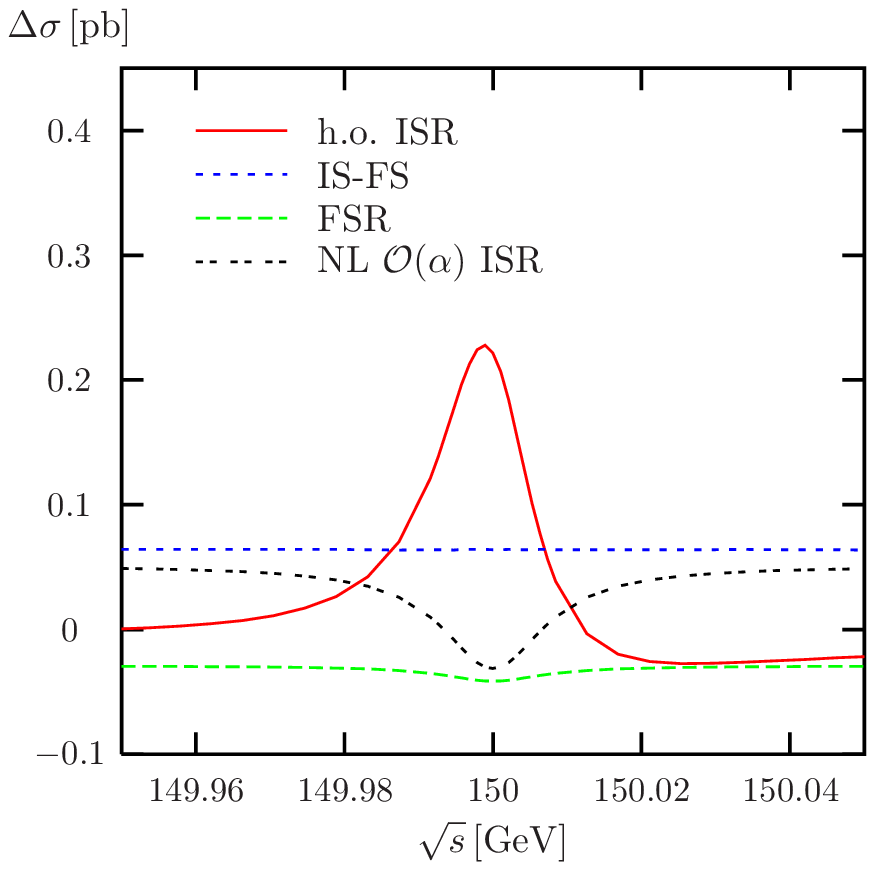}}
\end{picture} } 
\caption{Same as in \reffi{fig:mmbb_sm_mh115} but for $\MH=150\GeV$}
\label{fig:mmbb_sm_mh150}
\efi
On the l.h.s.\ the lowest-order predictions for the cross sections 
are successively improved by including the photonic and QCD corrections.
The invariant-mass cut \refeq{eq:Mhad} implies that the large
positive ISR correction of more than 100\% turns into a negative
correction of $\sim-4\%$ ($-8\%$) off resonance and of 
$\sim-49\%$ ($-46\%$) for the resonance part for $\MH=115\GeV$ ($150\GeV$).
This leads to a reduction of the resonance peak over the continuous
background.
The QCD corrections, on the other hand, tend to enhance the resonance
peak, since they are positive and larger on resonance than off resonance.
For $\MH=115\GeV$ ($150\GeV$) the resonant part of the cross section is 
enhanced by $\sim 21\%$ ($20\%$) compared to $\sim 5\%$ ($4\%$) in the 
continuum. This is in good agreement with the estimate 
that is based on the neglect of b-mass effects and of interferences
between the spin-1 and spin-0 channels.
In this approximation the results for the decays $\PZ,\PH\to\Pb\bar\Pb$
(see \citeres{LEP1,Braaten:1980yq})
can be used deduce the following behaviour of the QCD corrections,
\beq
\Delta\sigma_{\QCD}(s)\Big|_{\Mb\to0} \sim
\frac{\alpha_{\mathrm{s}}(\sqrt{s})}{\pi}\,\sigma_{\gamma/\PZ,0} +
\frac{17}{3}\frac{\alpha_{\mathrm{s}}(\sqrt{s})}{\pi}\,\sigma_{\PH,0},
\label{eq:QCDapprox}
\eeq
where $\sigma_{\gamma/\PZ,0}$ and $\sigma_{\PH,0}$ denote the 
contributions to the lowest-order cross section $\sigma_0$ induced by 
$\gamma/\PZ$ and Higgs-boson exchange, respectively.

The photonic corrections are dominated by the LL of the 
ISR corrections in $\O(\alpha)$. The remaining photonic 
contributions are shown on the r.h.s.\ of the figures. The largest
of these subleading corrections are due to the LL parts beyond
$\O(\alpha)$, followed by the non-leading (NL), i.e.\ non-logarithmic,
ISR contributions.
Note that the latter are the part of ISR that is not included in
the LL approach via the structure function convolution described
in \refse{se:isr}. The NL $\O(\alpha)$ corrections reduce the peak
over background by $\sim 4\%$ for both $\MH$ values.
The FSR and initial-final interaction effects turn out to be 
below $1\%$ of the continuous background and do not show sizable
variations in the resonance regions. Thus, they can be safely neglected
in all phenomenological investigations of the Higgs resonance.

The Higgs resonances inspected in \reffis{fig:mmbb_sm_mh115} and 
\ref{fig:mmbb_sm_mh150} are so narrow that the beam energy spread
of the muon collider will eventually determine the resonance shape
that is seen experimentally. As widely done in the literature,
we illustrate this effect by performing a Gaussian convolution
of the scattering cross section,
\beq
\bar\si\left(\sqrt{s}\right) =
\frac{1}{\sqrt{2\pi}\de_{\sqrt{s}}}
\int\rd\sqrt{\hat s}\; \si\left(\sqrt{\hat s}\right) \,
\exp\left\{-\frac{\left(\sqrt{\hat s}-\sqrt{s}\right)^2}
                 {2\de_{\sqrt{s}}^2}\right\},
\label{eq:smear}
\eeq
where $\sqrt{s}$ is the central value of the CM energy and 
$\de_{\sqrt{s}}$ the corresponding root-mean-square Gaussian spread.
Following \citere{Barger:2001mi} we use
\beq
\de_{\sqrt{s}} = 2\MeV \left(\frac{R}{0.003\%}\right)
\left(\frac{\sqrt{s}}{100\GeV}\right)
\eeq
with the actual value $R=0.003\%$. The results of applying this convolution
to the resonances shown in \reffis{fig:mmbb_sm_mh115} and 
\ref{fig:mmbb_sm_mh150} are shown in \reffis{fig:mmbb_sm_mh115_smear} and 
\ref{fig:mmbb_sm_mh150_smear}.
\bfi
\setlength{\unitlength}{1cm}
\centerline{
\begin{picture}(15.5,7.7)
\put(-5.0,-15.1){\includegraphics{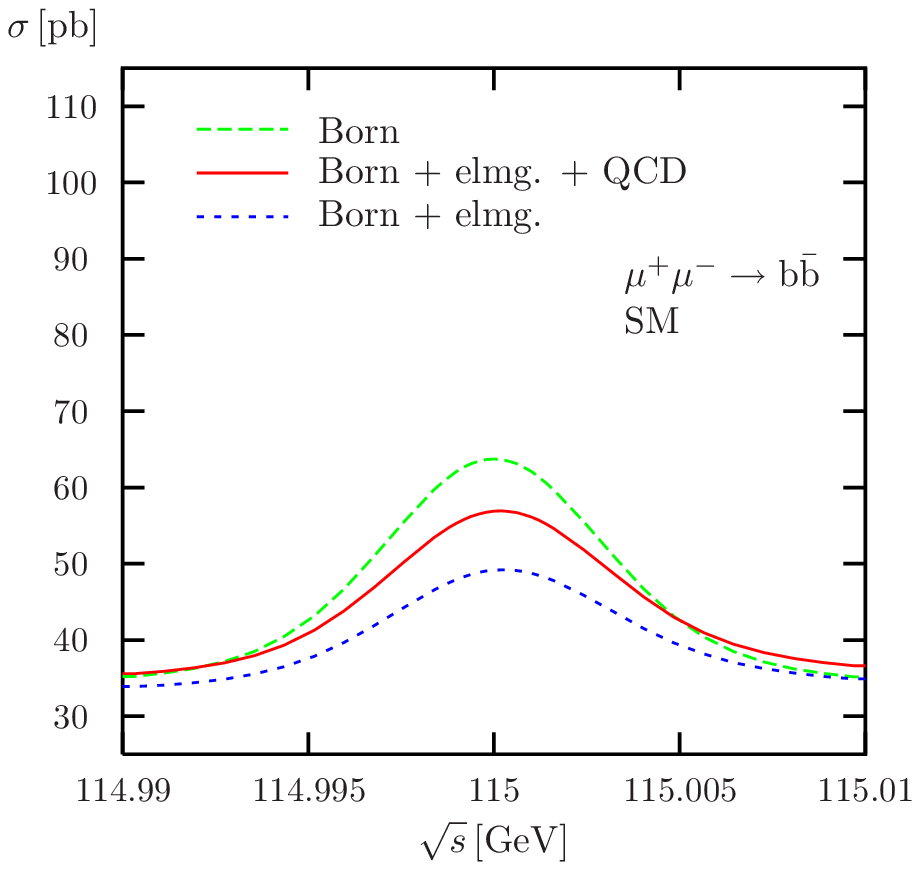}}
\put( 3.0,-15.1){\includegraphics{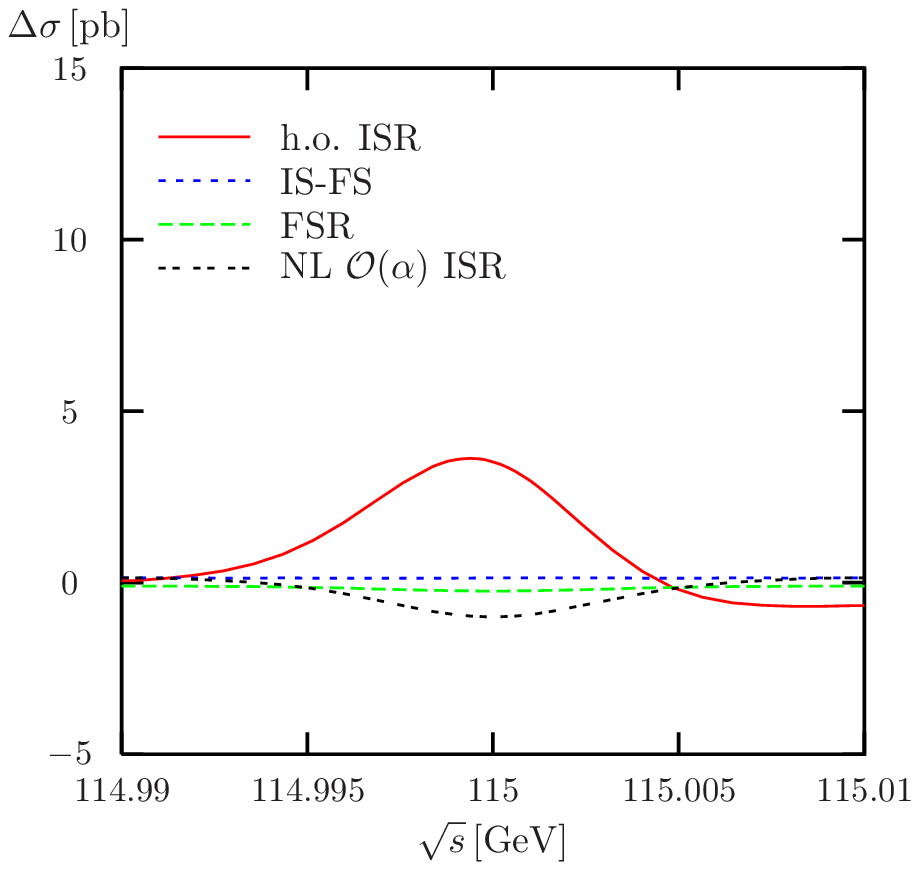}}
\end{picture} } 
\caption{SM cross section $\mu^-\mu^+\to\Pb\bar\Pb$ near the Higgs
resonance for $\MH=115\GeV$ (left) and some contributions
to the photonic corrections (right), as in \reffi{fig:mmbb_sm_mh115}, 
but now with beam energy smearing}
\label{fig:mmbb_sm_mh115_smear}
\vspace*{3em}
\centerline{
\begin{picture}(15.5,7.7)
\put(-5.0,-15.1){\includegraphics{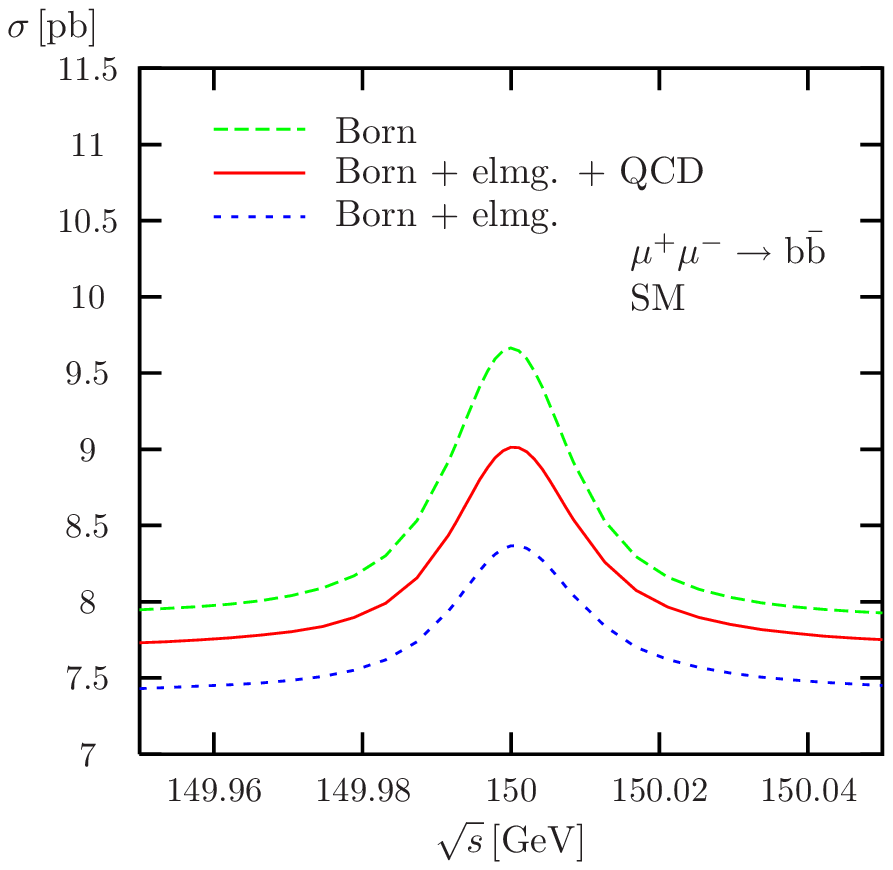}}
\put( 3.0,-15.1){\includegraphics{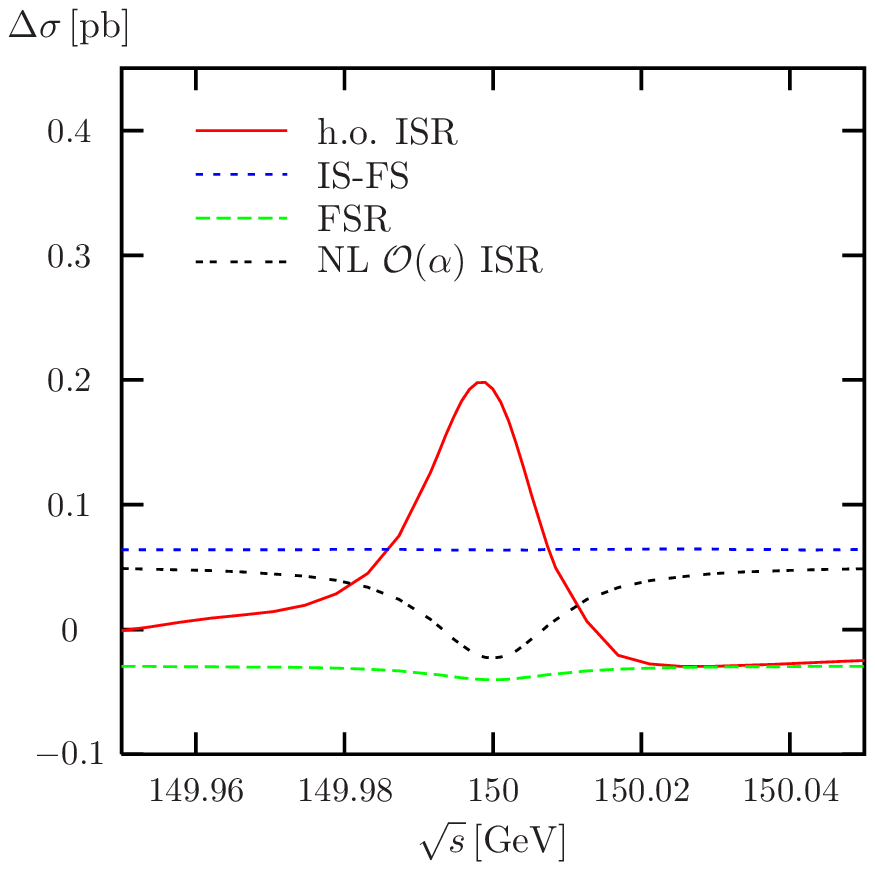}}
\end{picture} } 
\caption{Same as in \reffi{fig:mmbb_sm_mh115_smear} but for $\MH=150\GeV$}
\label{fig:mmbb_sm_mh150_smear}
\efi
As expected, the effect of smearing the resonance is much more pronounced
for $\MH=115\GeV$, where the ratio $\de_{\sqrt{s}}/\GH$ is 
$\sim 0.7$, than for $\MH=150\GeV$, where the ratio is $\sim 0.2$.
In both cases, however, the relative impact of the
various radiative corrections remains practically unchanged by the
smearing.

\subsection{\boldmath{The processes $\mu^-\mu^+\to\Pb\bar\Pb,\Pt\bar\Pt$
in the MSSM}}

Now we turn to Higgs production in the MSSM. Of course, it is beyond the
scope of this paper to give an exhaustive overview over the 
phenomenology of the full supersymmetric parameter space.
We rather concentrate on the two
MSSM parameter choices given in \refta{tab:GH}, which have been selected
to be still experimentally allowed and to cover the most interesting
patterns of Higgs resonances: 
For $\tan\be=30$ and $\MA=140\GeV$, the lightest Higgs boson h has
a mass of $\Mh\sim121\GeV$ with an intermediate width of $\Gh\sim 0.1\GeV$,
while the resonances of A and H practically lie on top of each other
at $\MH\sim\MA=140\GeV$ with large widths $\GH\sim\GA\sim 3\GeV$.
On the other hand, for $\tan\be=5$ and $\MA=400\GeV$, the h boson is
SM-like with a mass of $\Mh\sim116\GeV$ and a narrow width of $\Gh\sim4\MeV$;
the A and H resonances in this case are close to each other but
separated, since the mass difference $\MH-\MA\sim2\GeV$ is sufficiently
larger than the widths $\GH\sim0.5\GeV$ and $\GA\sim0.8\GeV$.
Note also that in the latter scenario the heavy Higgs bosons can decay
into $\Pt\bar\Pt$ pairs.
Since we do not consider the SM-like h~boson any further,
the widths of all resonances discussed in the following are
much larger than the achievable beam energy spread of a muon collider.
Therefore, the smearing hardly affects the resonance
shapes shown below, and we can omit the convolution \refeq{eq:smear}
in this section.

We first consider the scenario $\tan\be=30$ and $\MA=140\GeV$
in \reffis{fig:mmhbb_mssm_tgb30_ma140} and \ref{fig:mmHAbb_mssm_tgb30_ma140},
where the h and A/H resonances are shown together with the various
radiative corrections, in analogy to the previous section for the SM
Higgs boson. 
\bfi
\setlength{\unitlength}{1cm}
\centerline{
\begin{picture}(15.5,7.7)
\put(-5.0,-15.1){\includegraphics{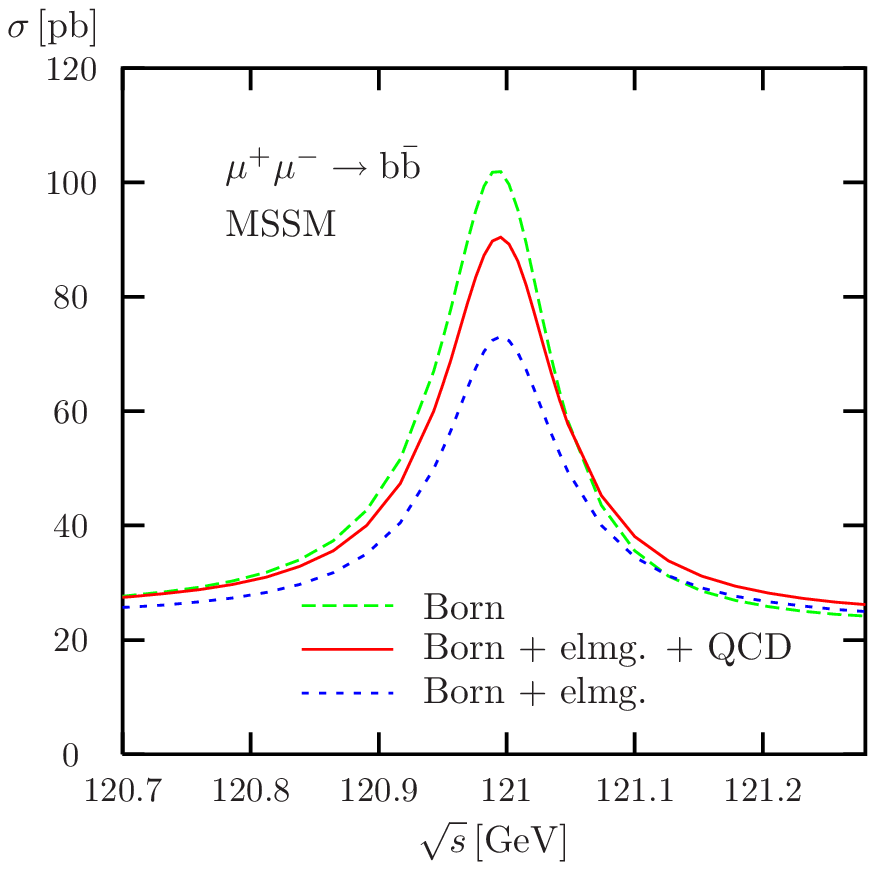}}
\put( 3.0,-15.1){\includegraphics{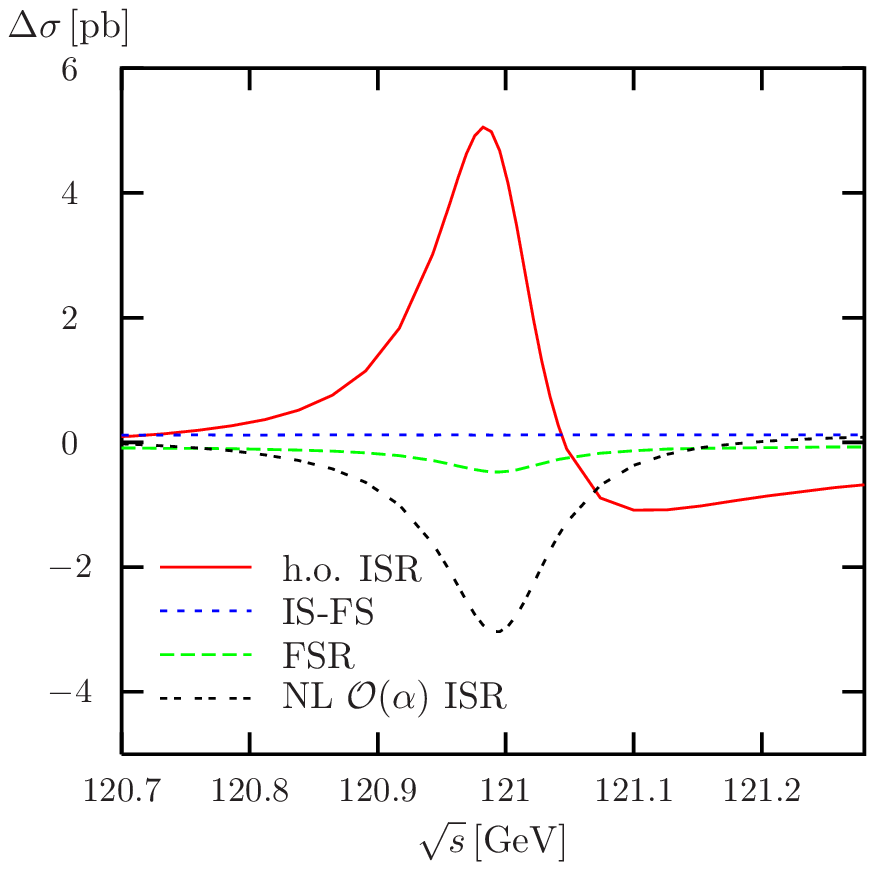}}
\end{picture} } 
\caption{MSSM cross section $\mu^-\mu^+\to\Pb\bar\Pb$ near the h
resonance for $\MA=140\GeV$ and $\tan\be=30$ (left) and some contributions
to the photonic corrections (right)}
\label{fig:mmhbb_mssm_tgb30_ma140}
\vspace*{3em}
\centerline{
\begin{picture}(15.5,7.7)
\put(-5.0,-15.1){\includegraphics{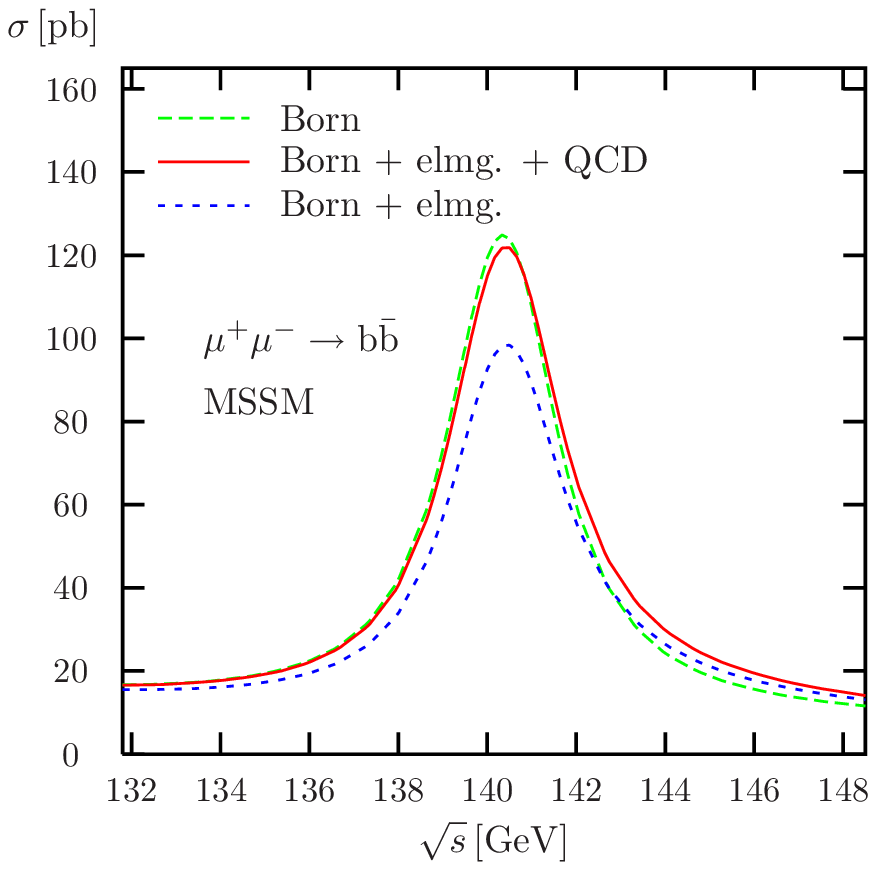}}
\put( 3.0,-15.1){\includegraphics{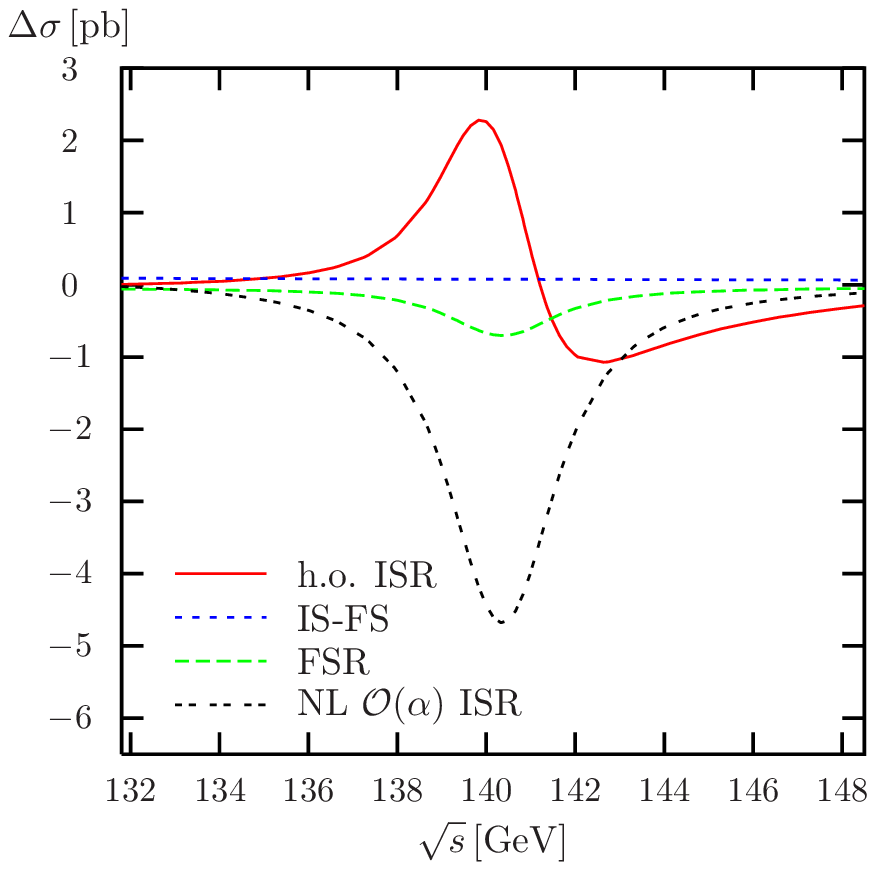}}
\end{picture} } 
\caption{MSSM cross section $\mu^-\mu^+\to\Pb\bar\Pb$ near the H and A
resonances (which lie on top of each other)
for $\MA=140\GeV$ and $\tan\be=30$ (left) and some contributions
to the photonic corrections (right)}
\label{fig:mmHAbb_mssm_tgb30_ma140}
\efi
The photonic corrections are dominated by the leading ISR logarithms,
and the QCD corrections follow the qualitative behaviour of the massless
approximation \refeq{eq:QCDapprox}, where $\sigma_{\PH,0}$ is replaced by
$\sigma_{\Ph/\PH/\PA,0}$.
Qualitatively the corrections are very similar to the
SM case, as expected. The ISR beyond $\O(\al)$ tends to enhance
the h and A/H resonances by $\sim 7\%$ and $\sim 2\%$, respectively,
but the NL ISR tends to reduce the resonances by $\sim 4\%$.
Photonic FSR corrections stay below $1\%$ everywhere, and initial-final
interferences do not show a resonance behaviour at all.

Of course, the corrections to the SM-like h-boson resonance of the second
scenario ($\tan\be=5$, $\MA=400\GeV$) also do not possess any new feature.
Therefore, this resonance is not shown explicitly here. 
More interestingly, \reffis{fig:mmHAbb_mssm_tgb5_ma400} and
\ref{fig:mmHAtt_mssm_tgb5_ma400} illustrate the A and H resonances,
which are close together, for the two processes
$\mu^+\mu^-\to\Pb\bar\Pb,\Pt\bar\Pt$.
\bfi
\setlength{\unitlength}{1cm}
\centerline{
\begin{picture}(15.5,7.7)
\put(-5.0,-15.1){\includegraphics{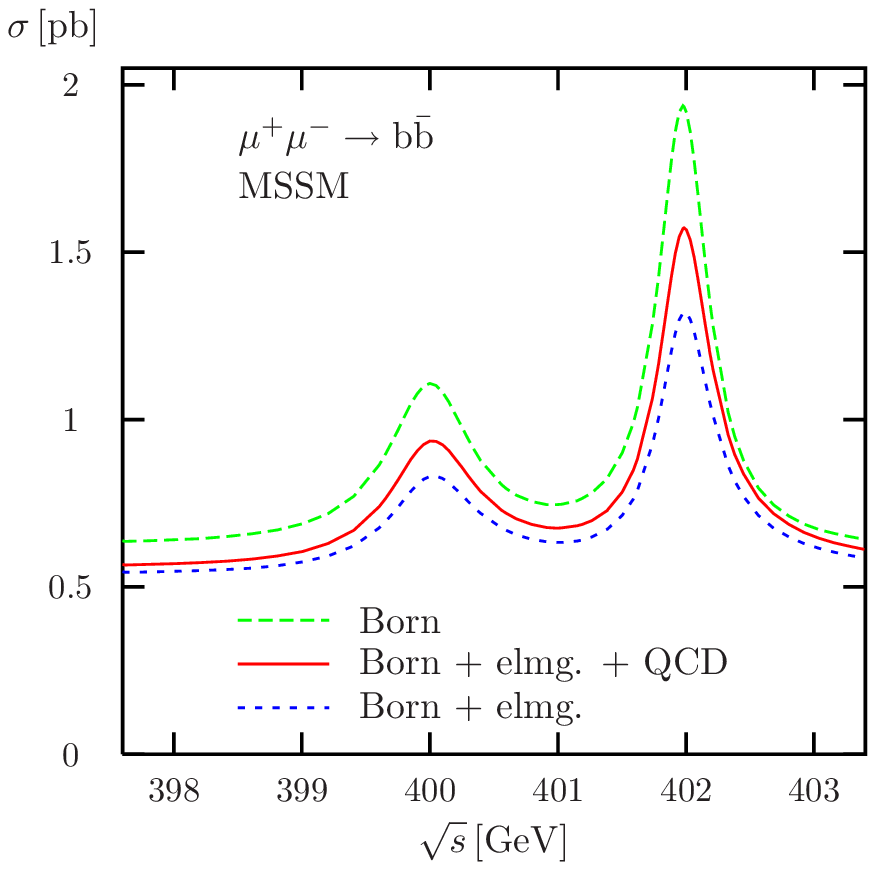}}
\put( 3.0,-15.1){\includegraphics{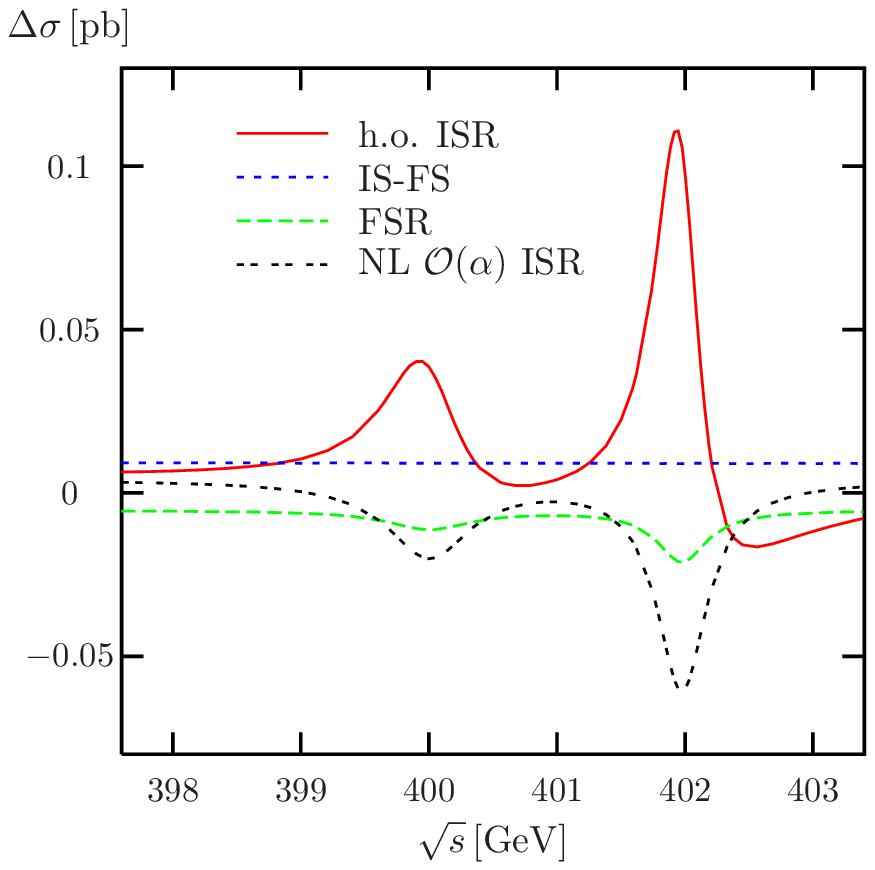}}
\end{picture} } 
\caption{MSSM cross section $\mu^-\mu^+\to\Pb\bar\Pb$ near the H and A
resonances for $\MA=400\GeV$ and $\tan\be=5$ (left) and some contributions
to the photonic corrections (right)}
\label{fig:mmHAbb_mssm_tgb5_ma400}
\vspace*{3em}
\centerline{
\begin{picture}(15.5,7.7)
\put(-5.0,-15.1){\includegraphics{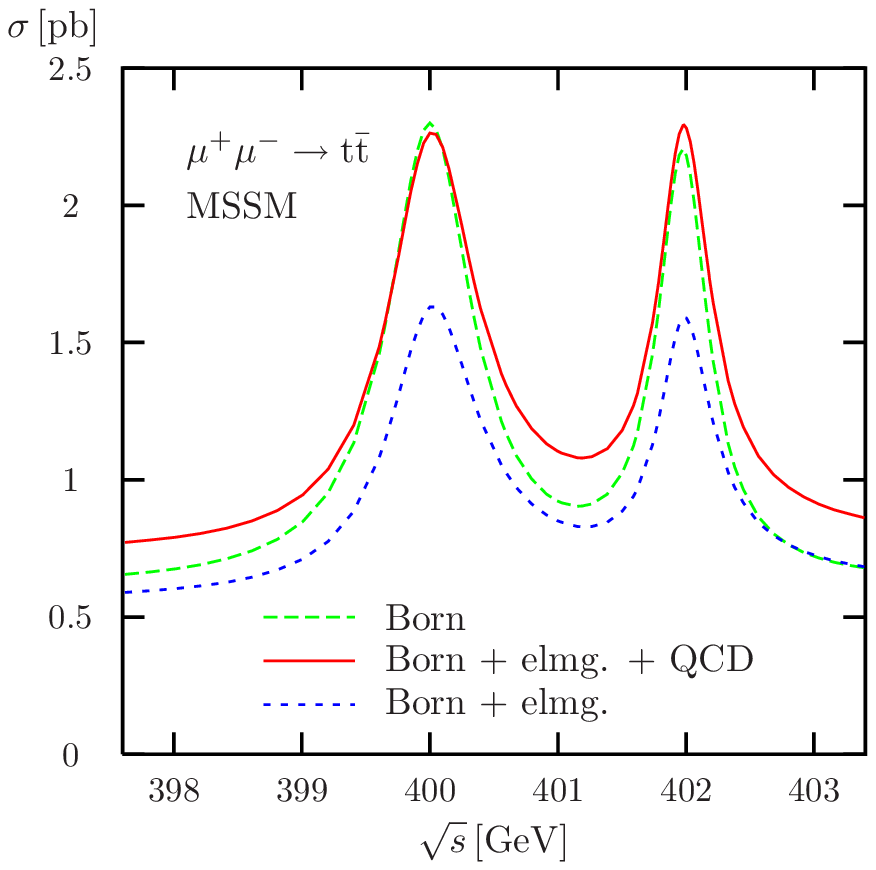}}
\put( 3.0,-15.1){\includegraphics{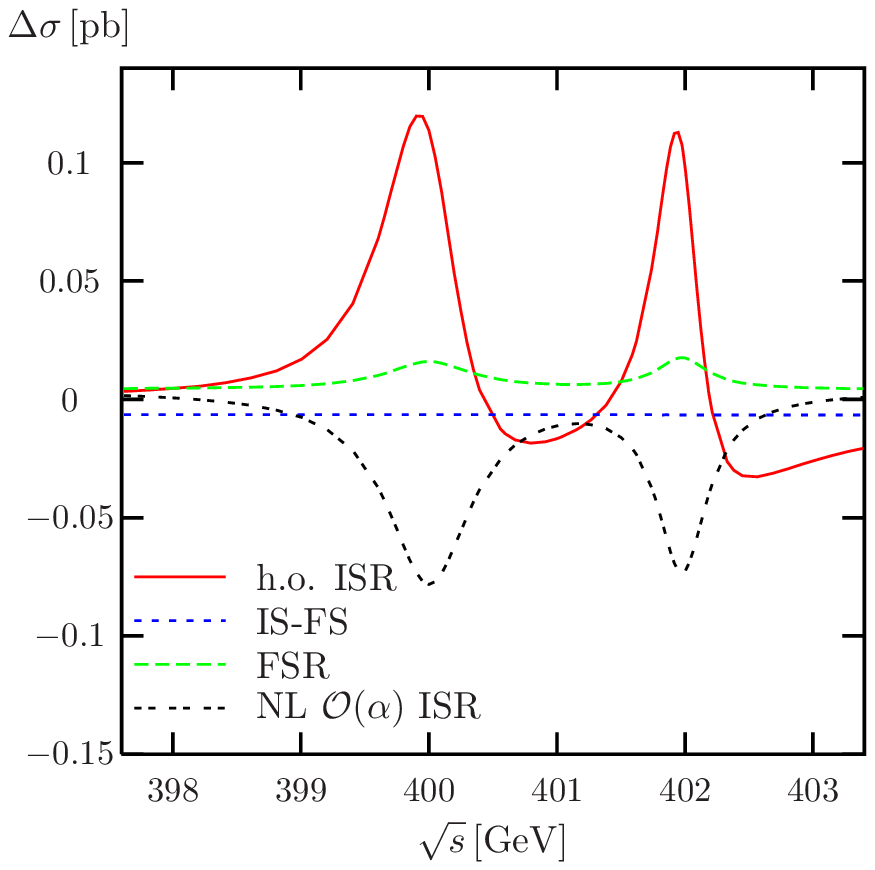}}
\end{picture} } 
\caption{MSSM cross section $\mu^-\mu^+\to\Pt\bar\Pt$ near the H and A
resonances for $\MA=400\GeV$ and $\tan\be=5$ (left) and some contributions
to the photonic corrections (right)}
\label{fig:mmHAtt_mssm_tgb5_ma400}
\efi
We note that we do not apply any cut like \refeq{eq:Mhad} in the
case of $\Pt\bar\Pt$ production, i.e.\ the total cross section is
considered, since a radiative return to the Z~resonance via ISR is not 
possible in this channel.
For each of the resonances in $\Pb\bar\Pb$ production the structure of 
the corrections again looks as in the previous cases, i.e.\ the
QCD corrections follow the pattern of Eq.~\refeq{eq:QCDapprox} and
the photonic corrections are dominated by the ISR logarithms.
The higher-order
ISR modifies the resonance cross section by $\sim 8\%$, NL ISR by $\sim -4\%$,
FSR by less than $1\%$, and initial-final interferences not at all.
For $\Pt\bar\Pt$ production only minor differences are observed.
The main differences is the failure of the approximation \refeq{eq:QCDapprox},
which is due to the large mass of the top quark.
The leading ISR effects still dominate the photonic corrections, while 
higher-order and NL ISR amount to $\sim 6\%$ and $\sim -5\%$ on resonance,
respectively. Owing to the omission of an invariant-mass cut on the
$\Pt\bar\Pt$ system, FSR now becomes positive, but nevertheless does not
exceed $\sim 1\%$, in spite of an enhancement by a factor 4 w.r.t.\
$\Pb\bar\Pb$ production induced by the quark charges. 
Finally, initial-final interferences again do not affect the
resonance shape.

\section{Conclusions}
\label{se:concl}

Muon colliders provide a unique framework for studying Higgs-boson
resonances, and thus for measuring Higgs masses, widths, and various
couplings. To this end, it is necessary to understand the resonance
shapes to high precision, i.e.\ in particular radiative
corrections have to be controlled.
In this paper the photonic and QCD radiative corrections to
$\mu^+\mu^-\to f\bar f$ have been discussed in detail, both for
the SM and the MSSM.
In the calculation the issue of a gauge-invariant description of the
resonances is emphasized, and the corrections that are relevant for 
the resonance shape are presented in analytical form.

The most important photonic corrections are due to ISR and 
amount to ${\cal O}(100\%)$.
Suppressing the radiative return to the Z-boson resonance
by appropriate invariant-mass cuts, the calculated ISR corrections,
which include the full ${\cal O}(\alpha)$ contributions and leading
logarithmic effects up to ${\cal O}(\alpha^3)$, should be sufficient
to describe photonic corrections within per-cent accuracy. 
Corrections due to photonic FSR modify the resonance by about $1\%$
or less. Initial-final interferences (box corrections) turn out
to be even smaller and non-resonant, i.e.\ they modify only the
continuous background of the resonance. In summary, photonic
corrections are now under control at the per-cent level.

QCD corrections have been calculated to ${\cal O}(\alpha_{\mathrm{s}})$
including the full mass dependence. Further improvements beyond
${\cal O}(\alpha_{\mathrm{s}})$ are certainly needed in the future.

The genuine weak corrections are not calculated in this paper. 
A consistent inclusion of these corrections to the full process
$\mu^+\mu^-\to f\bar f$ (keeping all fermion masses non-zero)
requires a gauge-invariant Dyson summation of all self-energies,
as discussed in some detail. 
However, for a theoretical description of the
Higgs resonances only, the relevant weak corrections can certainly
be incorporated in terms of effective couplings, simplifying the
task considerably.

\section*{Acknowledgements}

We thank A.~Denner, M.~Spira and P.M.~Zerwas for valuable discussions
and for carefully reading the manuscript.

\appendix
\setcounter{section}{1}

\section*{Appendix}

\section*{Standard matrix elements}

\newcommand{\MatVV}{\M^{vv}}
\newcommand{\MatAV}{\M^{av}}
\newcommand{\MatVA}{\M^{va}}
\newcommand{\MatAA}{\M^{aa}}
In this appendix we give explicit analytical expressions for the SME
introduced in Eq.~\refeq{eq:SME} of \refse{se:lo}, using again the WvdW
technique of \citere{Dittmaier:1999nn} as in
\refse{se:mmffa} for the bremsstrahlung matrix elements.
Inserting the expressions \refeq{eq:diracsp} for the Dirac spinors
into Eq.~\refeq{eq:SME}, the SME in terms of WvdW spinor products read
\beqar
\M^{xy}_{1} &=& 2\Big( \CPHIxi\phiXI + \de_x\Cpsixi\PSIXI
 + \de_y\CPHIETA\phieta + \de_x\de_y\CpsiETA\PSIeta \Big), \nn\\ 
\M^{xy}_{2} &=& \Big( \phiPSI + \de_x\CPHIpsi \Big)
                    \Big( \XIeta +\de_y\CxiETA \Big), \nn\\ 
\M^{xy}_{3} &=& \Big( \PHIqphi + \de_x\psiqPSI \Big)
                    \Big( \XIeta + \de_y\CxiETA \Big), \nn\\ 
\M^{xy}_{4} &=& \Big( \phiPSI + \de_x\CPHIpsi \Big)
                    \Big( \xipXI + \de_y\ETApeta \Big), 
\eeqar
where $x,y$ are the vector and axial-vector labels $v,a$ with the
sign factors
\beq
\de_v = +1, \qquad \de_a=-1.
\eeq
Moreover, we have used the shorthands
\beqar
\PHIqphi &=& \sum_{i=1,2}
\langle\phi'\rho_i\rangle^* \langle\phi\rho_i\rangle, \qquad
\psiqPSI  =  \sum_{i=1,2}
\langle\psi\rho_i\rangle^* \langle\psi'\rho_i\rangle, \nn\\
\ETApeta &=& \sum_{i=1,2} 
\langle\eta'\kappa_i\rangle^* \langle\eta\kappa_i\rangle, \qquad
\xipXI    =  \sum_{i=1,2}
\langle\xi\kappa_i\rangle^* \langle\xi'\kappa_i\rangle.
\eeqar
The explicit expressions for the various helicity configurations
are easily obtained by identifying the generic spinors $\phi$, $\psi$,
etc., with the appropriate insertions according to Eq.~\refeq{eq:wvdwins}.

\end{document}